\documentstyle[12pt,epsf]{article}
\newcommand{\be}{\begin{equation}}
\newcommand{\ee}{\end{equation}}

\newcommand{\beqa}{\begin{eqnarray}}
\newcommand{\eeqa}{\end{eqnarray}}
\newcommand{\nn}{\nonumber}

\newcommand{\eqref}[1]{(\ref{#1})}

\def\boxit#1{\vbox{\hrule\hbox{\vrule\kern8pt
\vbox{\hbox{\kern8pt}\hbox{\vbox{#1}}\hbox{\kern8pt}}
\kern8pt\vrule}\hrule}}
\def\mathboxit#1{\vbox{\hrule\hbox{\vrule\kern8pt\vbox{\kern8pt
\hbox{$\displaystyle #1$}\kern8pt}\kern8pt\vrule}\hrule}}

\def\IB{\relax\hbox{$\inbar\kern-.3em{\rm B}$}}
\def\IC{\relax\hbox{$\inbar\kern-.3em{\rm C}$}}
\def\ID{\relax\hbox{$\inbar\kern-.3em{\rm D}$}}
\def\IE{\relax\hbox{$\inbar\kern-.3em{\rm E}$}}
\def\IF{\relax\hbox{$\inbar\kern-.3em{\rm F}$}}
\def\IG{\relax\hbox{$\inbar\kern-.3em{\rm G}$}}
\def\IGa{\relax\hbox{${\rm I}\kern-.18em\Gamma$}}
\def\IH{\relax{\rm I\kern-.18em H}}
\def\IK{\relax{\rm I\kern-.18em K}}
\def\IL{\relax{\rm I\kern-.18em L}}
\def\IP{\relax{\rm I\kern-.18em P}}
\def\IR{\relax{\rm I\kern-.18em R}}
\def\IZ{\relax\ifmmode\mathchoice
{\hbox{\cmss Z\kern-.4em Z}}{\hbox{\cmss Z\kern-.4em Z}}
{\lower.9pt\hbox{\cmsss Z\kern-.4em Z}} {\lower1.2pt\hbox{\cmsss
Z\kern-.4em Z}}\else{\cmss Z\kern-.4em Z}\fi}

\def\II{\relax{\rm I\kern-.18em I}}

\begin{document}

\hfill  NRCPS-HE-05-44

\vspace{1cm}
\begin{center}
{\LARGE ~\\
Generalization of Yang-Mills Theory \footnote{\nn Dedicated to the 50th
anniversary of the  Yang-Mills theory\cite{yang}.}}\\
~\\
{\large Non-Abelian Tensor Gauge Fields\\
and\\
Higher-Spin Extension of Standard Model\\
~\\

}

\vspace{1cm}

{\sl George Savvidy\\
Demokritos National Research Center\\
Institute of Nuclear Physics\\
Ag. Paraskevi, GR-15310 Athens,Greece  \\
\centerline{\footnotesize\it E-mail: savvidy@inp.demokritos.gr}
}
\end{center}
\vspace{60pt}

\centerline{{\bf Abstract}}

\vspace{12pt}

\noindent
In the present paper we shall extend the gauge principle so that it will enlarge
the original algebra of the Abelian gauge transformations
found earlier in our studies of tensionless strings to the non-Abelian
case.  In this extension of the Yang-Mills theory the vector gauge boson
becomes a member of a bigger family of gauge bosons of arbitrary
large integer spins. The invariant Lagrangian does not
contain higher derivatives of tensor gauge fields
and  all interactions take place through three- and four-particle
exchanges with dimensionless coupling constant. The extended gauge
theory has the same index of divergences of its Feynman diagrams
as the Yang-Mills theory does and most probably will be renormalizable.
The proposed extension may lead  to a natural inclusion of the standard
theory of fundamental forces into a larger theory in which
vector gauge bosons, leptons and quarks represent a
low-spin subgroup of an enlarged family of particles with higher spins.
I analyze the masses of the new tensor gauge bosons, their decay and creation
processes in the extended standard model.


\newpage

\pagestyle{plain}

\section{{\it Introduction}}

It is well understood, that the concept of local gauge
invariance allows to define the non-Abelian gauge fields,
to derive their  dynamical field equations
and to develop a universal point of view on matter interactions
as resulting from the exchange of gauge quanta of different forms \cite{yang,yukawa}.
The fundamental forces - electromagnetic, weak and strong interactions
can successfully be described by the non-Abelian Yang-Mills fields.
The vector-like gauge particles - the photon, $W^{\pm},Z$ and
gluons mediate interaction
between smallest constituents of matter - leptons and quarks
\cite{Schwinger:1957em,Glashow:1961tr,salam,weinberg,Feynman:1963ax,Faddeev:1967fc,
DeWitt:1967ub,Mandelstam:1968hz,Slavnov:1970tk,'tHooft:1971fh,'tHooft:1971rn,'tHooft:1972fi,
Lee:1971kj,Lee:1972fj,Slavnov:1972fg,Taylor:1971ff,
Gross:1973id,Politzer:1973fx,Fritzsch:1973pi,Savvidy:1977as} .

The non-Abelian local gauge invariance, which was formulated
by Yang and Mills in\cite{yang},
requires that all interactions must be invariant under
independent rotations of internal
charges at all space-time
points\footnote{The early formulation of the Abelian gauge invariance
of the quantum electrodynamics was given in  \cite{fock,klein,london,weyl,pauli}
(see also \cite{chern}).}.
The gauge principle allows very little arbitrariness: the interaction of matter
fields,  which carry non-commuting internal charges, and the nonlinear
self-interaction of gauge bosons are essentially fixed by the requirement
of local gauge invariance, very similar to the self-interaction of
gravitons in general relativity.

It is therefore appealing to extend the gauge principle, which was elevated by Yang and
Mills to a powerful constructive principle \cite{yang}, so that it will define the
interaction of matter fields which carry
not only non-commutative internal charges, but
also arbitrary half-integer spins. It seems that this will naturally
lead to a theory in which fundamental forces will be mediated by
integer-spin gauge quanta  and
that the Yang-Mills vector gauge boson will become a member of a bigger
family of tensor gauge bosons.

Today there is no experimental evidence of the existence of such new particles
at the energy scale of order of few hundred GeV. But the standard string theory
predicts the existence of fundamental particles of arbitrary large spins
and masses of the order of the Planck mass scale, while the multiplicity of these
particles grows exponentially.

Alternatively, in the tensionless string theory with perimeter action,
the number of fundamental particles with large
spins grows linearly \cite{Savvidy:dv,Savvidy:2003fx}.
In this respect the number of states in the perimeter
model is much less compared with
the standard string theory  and is larger compared with the field
theory models of the Yang-Mills type. From this point of view it is therefore much
closer to the quantum field theory rather than to the standard string theory and
the main question which we would like to address in
this article is whether one can
formulate the corresponding field theory \cite{Savvidy:2003fx}.

{\it In the present paper we shall extend the gauge principle so that it will enlarge
the original algebra of the Abelian local gauge transformations
found in \cite{Savvidy:2003fx} to the non-Abelian case
and will allow to unify into one multiplet particles with arbitrary
large spins}. The proposed generalization may lead to a natural
inclusion of the standard
theory of fundamental forces into a larger theory in which standard
particles (vector gauge bosons, leptons and quarks) represent a
low-spin subgroup of an enlarged family of particles with higher spins.
The conjectured extension of the fundamental forces can in principle be
checked in future experiments.

It is important to note that the literature on higher-spin
fields is enormous and I shall not attempt to give a comprehensive
review of it here.
The early investigation of higher-spin representations of the Poincar\'e
algebra and of the corresponding field equations is due to Majorana,
Dirac and Wigner\cite{majorana,dirac,wigner}. The theory of massive particles
of higher spin was further developed by Fierz and Pauli \cite{fierzpauli} and
Rarita and Schwinger \cite{rarita}. The Lagrangian and
S-matrix formulations of {\it free field
theory} of massive and massless fields with higher spin
have been completely constructed in
\cite{schwinger,Weinberg:1964cn,Weinberg:1964ev,Weinberg:1964ew,
chang,singh,singh1,fronsdal,fronsdal1}.

The problem of {\it introducing interaction} appears to be much more complex.
There is a large amount of publications
devoted to the self-interaction of higher-spin fields
\cite{Gupta,kraichnan,thirring,feynman,deser,fronsdal2,Sagnotti:2005ns}.
The main idea is to introduce self-interaction using iterations:
starting from the free quadratic Lagrangian for the higher-spin field
one should introduce a cubic, quadratic and higher-order
terms to the free Lagrangian
and then check, whether the so deformed algebra of the initial
group of transformations (\ref{symmetricgaugetransformation})
still forms a closed algebraic structure.
This program was able to reproduce successfully general relativity, but
met enormous difficulties for spin fields higher than two
\cite{witten,deser1,berends,dewit}. There is also
important development of interacting field theories in anti-de Sitter
space-time background, which is
reviewed in \cite{vasiliev,Sezgin:2001zs}.

The first positive result in this direction was
the light-front construction of the cubic
interaction term for the massless field of helicity $\pm \lambda$ in
\cite{Bengtsson:1983pd,Bengtsson:1983pg}.
As it was mentioned by the authors, it was important to
generalize their construction to higher-order vertices
in order to see, whether the theory is
fully consistent. They also stressed that there are $\lambda$
derivatives in the cubic interaction
vertex, the coupling constant has dimension of $[mass]^{1 -\lambda}$ and
that the dimensional coupling constant would be difficult to
handle in quantum theory because the index of divergence
$r=\lambda -1$ is not equal to zero.
This result of Lars Brink and his collaborators
raised expectations that a consistent interacting
theory might exist in flat space-time.

In our approach the gauge fields are defined as rank-$(s+1)$ tensors
$$
A^{a}_{\mu\lambda_1 ... \lambda_{s}}(x),~~~~~s=0,1,2,...
$$
and are totally symmetric with respect to the
indices $  \lambda_1 ... \lambda_{s}  $.  A priory the tensor fields
have no symmetries with
respect to the first index  $\mu$. This is an essential departure from the
previous considerations, in which the higher-rank tensors were totally symmetric
\cite{fierzpauli,schwinger,singh,fronsdal}.
The index $s$ runs from zero to infinity.
The first member of this family of the tensor gauge bosons is the Yang-Mills
vector boson $A^{a}_{\mu}$.

The fermions are defined as Rarita-Schwinger tensor-spinors \cite{rarita}
$$
\psi^{\alpha}_{\lambda_1 ... \lambda_{s}}(x)
$$
with mixed transformation properties of Dirac four-component wave
function (the index $\alpha$ denotes the Dirac index) and
are totally symmetric tensors of the rank $s$
over the indices $\lambda_1 ... \lambda_{s}$.
All fields of the $\{ \psi \}$ family
are isotopic multiplets belonging to the
same representation $\sigma$ of the semisimple Lie group G
(the corresponding indices are suppressed).
The bosonic matter is defined as totally symmetric Fierz-Pauli rank-s tensors
\cite{fierzpauli}
$$
\phi_{\lambda_1 ... \lambda_{s}}(x)
$$
all belonging  to the
same representation $\tau$ of the semisimple Lie group G.

We shall enlarge the global transformation group of
the matter fields in a way which has been
found in \cite{Savvidy:2003fx}
(expression (64) in \cite{Savvidy:2003fx}).
The extended isotopic transformation
of the fermion fields is defined as (\ref{mattertransformation})
\beqa
\delta_{\xi}  \psi(x) &=& -i \xi(x) ~ \psi(x),\nn\\
\delta_{\xi}  \psi_{\lambda_1}(x) &=& -i \xi(x) ~ \psi_{\lambda_1}(x),
-i  \xi_{\lambda_1}(x) ~\psi(x)\nn\\
.............&.&.....................................,\nn
\eeqa
where the infinitesimal group parameters $\xi_{\lambda_{1} ...\lambda_s }(x)=
\sigma_{a} \xi^{a}_{\lambda_{1} ...\lambda_s }(x) $ are
totally symmetric rank-s tensors and are the matrices of the corresponding representation
$\sigma^{ij}_{a}$ of the semisimple Lie group G. As one can check,
the extended isotopic transformations form a closed  algebraic structure:
$$[~\delta_{\eta},\delta_{\xi}]~\psi_{\lambda_1\lambda_2 ...\lambda_s} =
i~ \delta_{\zeta} \psi_{\lambda_1\lambda_2 ...\lambda_s}.
$$
The transformation of the bosonic matter fields $
\phi_{\lambda_1 ... \lambda_{s}}(x)$ is defined in the same way
(\ref{scalartransformation})\footnote{The representations can
be different for $\{\psi\}$ and $\{\phi\}$ families.}.

The extended non-Abelian gauge transformations of the tensor gauge fields are defined
by the following equations (\ref{polygauge}), (\ref{matrixform}),
(\ref{generalgaugetransform}):
\beqa\label{symmetryintroduction}
\delta_{\xi}  A_{\mu} &=& \partial_{\mu}\xi -i g[A_{\mu},\xi] ,\nonumber\\
\delta_{\xi}  A_{\mu\lambda_1} &=& \partial_{\mu}\xi_{\lambda_1} -
i g[A_{\mu},\xi_{\lambda_1}]
-i g [A_{\mu\lambda_1},\xi], \nonumber\\
&~&..............................,\nn
\eeqa
where the tensor gauge fields are in the matrix representation
$A^{ab}_{\mu\lambda_1 ... \lambda_{s}} =
(L_c)^{ab}  A^{c}_{\mu\lambda_1 ... \lambda_{s}} = i f^{acb}A^{c}_{\mu
\lambda_1 ... \lambda_{s}}$,
and  $L^a$ are the generators of the semisimple Lie group G in the adjoint representation.
Because the infinitesimal gauge parameters $\xi^{b}_{\lambda_{1} ...\lambda_s }$ are
totally symmetric rank-s tensors, the extended gauge transformation
leaves the symmetries of the tensor fields intact. These extended gauge transformations
generate a closed algebraic structure. To see that, one should compute the
commutator of two extended gauge transformations $\delta_{\eta}$ and $\delta_{\xi}$
of parameters $\eta$ and $\xi$.
The commutator of two transformations can be expressed in the form
$$
[~\delta_{\eta},\delta_{\xi}]~A_{\mu\lambda_1\lambda_2 ...\lambda_s} ~=~
-i g~ \delta_{\zeta} A_{\mu\lambda_1\lambda_2 ...\lambda_s}
$$
and is again an extended gauge transformation with the gauge parameters
$\{\zeta\}$ which are given by the matrix commutators (\ref{commutatorofparameterslow})
(\ref{generalalgebralow})
\beqa
\zeta&=&[\eta,\xi]\nn\\
\zeta_{\lambda_1}&=&[\eta,\xi_{\lambda_1}] +[\eta_{\lambda_1},\xi]\nn\\
......&.&..........................\nn
\eeqa
The first three terms of the invariant Lagrangian have the following form
(\ref{secondranklagrangian}), (\ref{fulllagrangian1}), (\ref{fulllagrangian2}):
$$
{{\cal L}} =  {{\cal L}}_1 +  g_2 {{\cal L}}_2 +... =-{1\over 4}G^{a}_{\mu\nu}
G^{a}_{\mu\nu} + g_2 \{
-{1\over 4}G^{a}_{\mu\nu,\lambda}G^{a}_{\mu\nu,\lambda}
-{1\over 4}G^{a}_{\mu\nu}G^{a}_{\mu\nu,\lambda\lambda}~\}~+~...,
$$
where the first term is the Yang-Mills Lagrangian, the
second and the third ones describe the tensor gauge fields $A^{a}_{\mu\nu},
A^{a}_{\mu\nu\lambda}$ and $g_2$ is the new coupling constant.
The generalized field strengths in this expression
are defined as (\ref{fieldstrength2}),(\ref{fieldstrength3}):
\beqa
G^{a}_{\mu\nu,\lambda} &=&
\partial_{\mu} A^{a}_{\nu\lambda} - \partial_{\nu} A^{a}_{\mu\lambda} +
g f^{abc}(~A^{b}_{\mu}~A^{c}_{\nu\lambda} + A^{b}_{\mu\lambda}~A^{c}_{\nu} ~),\nn\\
G^{a}_{\mu\nu,\lambda\rho} &=&
\partial_{\mu} A^{a}_{\nu\lambda\rho} - \partial_{\nu} A^{a}_{\mu\lambda\rho} +
g f^{abc}(~A^{b}_{\mu}~A^{c}_{\nu\lambda\rho} +
 A^{b}_{\mu\lambda}~A^{c}_{\nu\rho}+A^{b}_{\mu\rho}~A^{c}_{\nu\lambda}
 + A^{b}_{\mu\lambda\rho}~A^{c}_{\nu} ~),\nn\\
 ......&.&............................................\nn
\eeqa
and transform homogeneously with respect to the extended
gauge transformations. The field strength tensors are
antisymmetric in their first two indices and are totaly symmetric with respect to the
rest of the indices. By induction the entire construction
can be generalized to include tensor fields of any rank s.

It is important that:  i) {\it the Lagrangian does not
contain higher derivatives of tensor gauge fields
and that ii) all interactions take place
through the three- and four-particle exchanges with dimensionless
coupling constant iii) the extended gauge transformations
do not define the coupling constant $g_{2}$}.

The coupling constant $g_{2}$ remains arbitrary because
every term in the sum is separately gauge invariant and
the extended gauge symmetry alone does not define it.
The immediate conclusion which
can be drawn from above properties is that the extended gauge theory will have
the same index of divergences of its Feynman diagrams as the Yang-Mills theory
does and most probably will be renormalizable. The other conclusion is that
the second-rank tensor gauge field $A^{a}_{\mu\nu}$, which in our theory is an arbitrary
nonsymmetric tensor $A^{a}_{\mu\nu} \neq A^{a}_{\nu\mu}$,
does not directly coincide with the graviton,
because it is a charged gauge field and it has different gauge
symmetries and interactions.

As a next step we shall  introduce the gauge invariant interaction of
the tensor gauge bosons with the higher-rank fermion
and boson matter fields. The invariant Lagrangian for fermions
has the form (\ref{fermionlagrangian})
$$
{{\cal L}}_{F} = \bar{\psi}  ( i \not\!\partial +
g \not\!\!A )\psi+
\bar{\psi}_{\lambda}   ( i\not\!\partial + g\not\!\!A  )\psi_{\lambda} +....
$$
and for bosons (\ref{scalarlagrangian})
$$
{{\cal L}}_{B} =
\nabla_{\mu}\phi^{+} ~ \nabla_{\mu}\phi +
\nabla_{\mu}\phi^{+}_{\lambda} ~\nabla_{\mu}\phi_{\lambda}
+g^2  \phi^{+} A_{\mu\lambda} A_{\mu\lambda}\phi +....
$$

The proposed gauge invariant theory of interacting higher-spin particles
leads  to a natural inclusion of the standard
model of fundamental forces into a larger theory in which standard
particles (vector gauge bosons, leptons and quarks) represent a
low-spin subgroup of an enlarged family of particles with higher spins.
The conjectured extension of the fundamental forces can in principle be
checked in future experiments. This possibility is explored in the following
sections.

In the second section we shall outline the derivation of the pure Yang-Mills
theory, the transformation properties of the gauge fields, the definition of
the corresponding field stress tensor and expression for the invariant Lagrangian.
Necessary matrix notations are also introduced.

In the third section we shall introduce the tensor gauge fields of the second rank
together with their extended gauge transformation properties and shall prove that
these transformations form a closed algebraic structure.
The corresponding generalization of the field stress tensor will also be defined.
We shall present the invariant
Lagrangian for the second-rank tensor gauge fields and derive their classical
field equations. In section four this construction will be extended
to include tensor gauge fields of the third rank. In the fifth section we shall present
the general construction for the tensor gauge fields of arbitrary rank s.

In the sixth section we shall incorporate into the theory the fermions of
half-integer spins.  We shall construct the  gauge
invariant Lagrangian describing interaction of the tensor gauge
fields with half-integer spin fermions. In the seventh section
this construction will be
extended to include scalar fields and integer-spin  bosonic matter.
The Higgs mechanism will be extended to generate masses of the tensor gauge bosons.

In the eighth and ninth sections a tensor extension of the strong and electroweak
theories will be suggested.
I analyze the masses of the new tensor gauge bosons, their decay
and creation processes in the extended standard model.
Section ten contains the generalization of the non-Abelian transformations to
totally symmetric tensor gauge fields.
Section eleven contains some concluding remarks.

\section{{\it Yang-Mills Fields }}

The dynamics of the vector potential describing
spin-1 Yang-Mills quanta $A^{a}_{\mu}$ is given by the Lagrangian \cite{yang}:
\be
{{\cal L}}_1 = -{1\over 4}G^{a}_{\mu\nu}G^{a}_{\mu\nu},
\ee
where
\be\label{gemunu}
G^{a}_{\mu\nu} =
\partial_{\mu} A^{a}_{\nu} - \partial_{\nu} A^{a}_{\mu} +
g f^{abc}~A^{b}_{\mu}~A^{c}_{\nu}.
\ee
It is invariant with respect to the non-Abelian gauge transformations:
\be\label{standardgaugetransformation}
\delta A^{a}_{\mu} = ( \delta^{ab}\partial_{\mu}
+g f^{acb}A^{c}_{\mu})\xi^b .
\ee
The invariance of the Lagrangian ${{\cal L}}_1$ can be
demonstrated if one derives the transformation property of the
field strength $G^{a}_{\mu\nu}$ induced by the transformation
law (\ref{standardgaugetransformation}) of the gauge potential $A^{a}_{\mu}$.
Indeed, using definition
of $G^{a}_{\mu\nu}$ in (\ref{gemunu}) and the property of the
structure constant $f^{abc}$ one can get
$$
\delta G^{a}_{\mu\nu}= g f^{abc} G^{b}_{\mu\nu} \xi^c
$$
and therefore
$$
\delta {{\cal L}}_1 = -{1\over 2} G^{a}_{\mu\nu}~ \delta G^{a}_{\mu\nu} =
-{1\over 2} G^{a}_{\mu\nu}~ g f^{abc} G^{b}_{\mu\nu} \xi^c =0.
$$
It is also useful to introduce matrix representation for the gauge
fields $A^{ab}_{\mu} = A^{c}_{\mu} (L^c)_{ab} = i f^{acb}A^{c}_{\mu}$,
where $L^a$ are the generators of the semisimple Lie group G :
$$
[L^a , L^b] = i f^{abc} L^c ,
$$
where the structure constants fulfil the Jacobi identity $f^{aed}f^{bdc}
+ f^{bed}f^{cda} + f^{adb}f^{ced} =0$. The unitary transformation matrix
is equal to $U(\xi) = exp(ig \xi )$, where $\xi = L^{a} \xi^{a}$.
In the adjoint representation $(L^b)_{ac} = i f_{abc}$, and the
covariant derivative will take the form
$$
\nabla^{ab}_{\mu} = (\partial_{\mu}-ig A_{\mu})^{ab}.
$$
The gauge transformation can be represented also in a matrix form
$$
\delta A_{\mu} = \partial_{\mu}\xi -ig [A_{\mu},\xi],
$$
as well as the field strength
$$
G_{\mu\nu} = \partial_{\mu} A_{\nu} - \partial_{\nu} A_{\mu} -i g [ A_{\mu}~A_{\nu}].
$$
Finally the classical field equations are:
$$
\nabla^{ab}_{\mu} G^{b}_{\mu\nu} =0.
$$

\section{\it Second-Rank Tensor Gauge Fields}
In order to describe rank-two quanta we shall
introduce tensor gauge field $A^{a}_{\mu\nu}$
together with the higher-tensor gauge field $A^{a}_{\mu\nu\lambda}$, which is symmetric
with respect to the last two indices  ~~
$A^{a}_{\mu\nu\lambda} =A^{a}_{\mu\lambda\nu}$.
Using these fields we shall define the Lagrangian as
\be\label{secondranklagrangian}
{{\cal L}}_2 = -{1\over 4}G^{a}_{\mu\nu,\lambda}G^{a}_{\mu\nu,\lambda}
-{1\over 4}G^{a}_{\mu\nu}G^{a}_{\mu\nu,\lambda\lambda}~~,
\ee
where I have introduced the corresponding new field strengths of the form
\be\label{fieldstrength2}
G^{a}_{\mu\nu,\lambda} =
\partial_{\mu} A^{a}_{\nu\lambda} - \partial_{\nu} A^{a}_{\mu\lambda} +
g f^{abc}(~A^{b}_{\mu}~A^{c}_{\nu\lambda} + A^{b}_{\mu\lambda}~A^{c}_{\nu} ~)
\ee
and
\be\label{fieldstrength3}
G^{a}_{\mu\nu,\lambda\rho} =
\partial_{\mu} A^{a}_{\nu\lambda\rho} - \partial_{\nu} A^{a}_{\mu\lambda\rho} +
g f^{abc}(~A^{b}_{\mu}~A^{c}_{\nu\lambda\rho} +
 A^{b}_{\mu\lambda}~A^{c}_{\nu\rho}+A^{b}_{\mu\rho}~A^{c}_{\nu\lambda}
 + A^{b}_{\mu\lambda\rho}~A^{c}_{\nu} ~).
\ee
As one can see, the extended field strength tensors (\ref{fieldstrength2}) and
(\ref{fieldstrength3}) contain not only
tensor gauge fields of the corresponding order, but also
lower-rank tensor gauge fields.  The field strength $G^{a}_{\mu\nu,\lambda}$
contains in addition to the tensor field $A^{a}_{\mu\nu}$
the vector field $A^{a}_{\mu}$.
The field strength $G^{a}_{\mu\nu,\lambda\rho}$ contains
in addition to the tensor field $A^{a}_{\mu\nu\lambda}$ the
vector field $A^{a}_{\mu}$ and the second-rank tensor gauge field $A^{a}_{\mu\nu}$.
It is obvious from these expressions that without lower-rank tensor gauge
fields it would be impossible to construct the above
higher-rank field strength tensors. The extended field strength tensors are
antisymmetric in their first two indices and are totaly symmetric with respect to the
rest of the indices:
\beqa
G^{a}_{\mu\nu,\lambda} = -G^{a}_{\nu\mu,\lambda},~~~~
G^{a}_{\mu\nu,\lambda\rho} = -G^{a}_{\nu\mu,\lambda\rho},~~~
G^{a}_{\mu\nu,\lambda\rho} = G^{a}_{\mu\nu,\rho\lambda} ~.
\eeqa
One can also represent the field strengths in a matrix form:
\beqa
G_{\mu\nu,\lambda} &=&
\nabla_{\mu} A_{\nu\lambda} - \nabla_{\nu} A_{\mu\lambda}~~~,\nn\\
G_{\mu\nu,\lambda\rho} &=&
\nabla_{\mu} A_{\nu\lambda\rho} - \nabla_{\nu} A_{\mu\lambda\rho} -i g
[ A_{\mu\lambda}~A_{\nu\rho}] -i g [A_{\mu\rho}~A_{\nu\lambda}].
\eeqa

We have to prove that the Lagrangian ${{\cal L}}_2$ is invariant under
extended gauge transformations which I shall define as follows:
\beqa\label{polygauge}
\delta A^{a}_{\mu} &=& ( \delta^{ab}\partial_{\mu}
+g f^{acb}A^{c}_{\mu})\xi^b ,~~~~~ \nonumber\\
\delta A^{a}_{\mu\nu} &=&  ( \delta^{ab}\partial_{\mu}
+  g f^{acb}A^{c}_{\mu})\xi^{b}_{\nu} + g f^{acb}A^{c}_{\mu\nu}\xi^{b},\nonumber\\
\delta A^{a}_{\mu\nu \lambda}& =&  ( \delta^{ab}\partial_{\mu}
+g f^{acb} A^{c}_{\mu})\xi^{b}_{\nu\lambda} +
g f^{acb}(  A^{c}_{\mu  \nu}\xi^{b}_{\lambda } +
A^{c}_{\mu \lambda }\xi^{b}_{ \nu}+
A^{c}_{\mu\nu\lambda}\xi^{b}).
\eeqa
In addition to the standard gauge parameter $\xi^{a}$ in this
transformation law we have introduced new higher-rank gauge parameters:
the vector $\xi^{a}_{\mu}$ and the totally symmetric second-rank
tensor $\xi^{a}_{\mu\nu}$. This transformation respects the symmetry
properties of
the third rank gauge field $A_{\mu\nu\lambda} = A_{\mu\lambda\nu}$
\footnote{The extension of
the gauge transformations (\ref{polygauge}) to {\it totally symmetric}
higher-rank tensor gauge fields is presented in the tenth section.}.

First let us prove that a commutator of two of extended gauge transformations
can be expressed as a similar gauge transformation, and therefore extended gauge
transformations (\ref{polygauge}) form a closed algebraic structure.
To make the calculation more transparent let us
express the transformation law (\ref{polygauge}) in a matrix form:
\beqa\label{matrixform}
\delta_{\xi}  A_{\mu} &=& \partial_{\mu}\xi -i g[A_{\mu},\xi]\nonumber\\
\delta_{\xi}  A_{\mu\nu} &=& \partial_{\mu}\xi_{\nu} -i g[A_{\mu},\xi_{\nu}]
-i g [A_{\mu\nu},\xi]\nonumber\\
\delta_{\xi}  A_{\mu\nu\lambda} &=& \partial_{\mu}\xi_{\nu\lambda}
-i g[A_{\mu},\xi_{\nu\lambda}]-
i g[A_{\mu\nu},\xi_{\lambda}]-i g [A_{\mu\lambda},\xi_{\nu}]
-i g [A_{\mu\nu\lambda},\xi],
\eeqa
where $A_{\mu\nu} = A^{a}_{\mu\nu} L^a$, $A_{\mu\nu\lambda} =
A^{a}_{\mu\nu\lambda} L^a$ and $\xi = L^{a} \xi^{a}$.
The commutator of two gauge transformations acting
on a second-rank tensor gauge field is:
\beqa
[\delta_{\eta},\delta_{\xi}]A_{\mu\nu} &=& \delta_{\eta}~(-i g[A_{\mu},\xi_{\nu}]
-i g [A_{\mu\nu},\xi])-\nonumber\\
&-& \delta_{\xi}~(-i g[A_{\mu},\eta_{\nu}]
-i g [A_{\mu\nu},\eta])\nonumber\\
&=&-ig~\{~\partial_{\mu}([\eta,\xi_{\nu}] +[\eta_{\nu},\xi]) -i
g[A_{\mu},([\eta,\xi_{\nu}] +[\eta_{\nu},\xi])]
-i g[A_{\mu\nu},[\eta,\xi]] ~\}\nonumber\\
&=& -ig~\{~\partial_{\mu}\zeta_{\nu} -i g[A_{\mu},\zeta_{\nu}]
-i g [A_{\mu\nu},\zeta]~\} = -i g ~\delta_{\zeta}A_{\mu\nu}\nn
\eeqa
and is again a gauge transformation with gauge parameters
$\zeta^a, \zeta^{a}_{\mu}$ which are given by the following expressions:
$$
\zeta =[\eta,\xi],~~~~~~~\zeta_{\nu} = [\eta,\xi_{\nu}] +[\eta_{\nu},\xi].
$$
The commutator of two gauge transformations acting on a rank-3 tensor gauge field is:
\beqa
[\delta_{\eta},\delta_{\xi}]A_{\mu\nu\lambda} &=&
\delta_{\eta}~(-i g[A_{\mu},\xi_{\nu\lambda}]
-i g  [A_{\mu\nu},\xi_{\lambda}]-i g  [A_{\mu\lambda},\xi_{\nu}]-
i g [A_{\mu\nu\lambda},\xi]) \nn\\
&-& \delta_{\xi}~(-i g[A_{\mu},\eta_{\nu\lambda}]
-i g  [A_{\mu\nu},\eta_{\lambda}]-i g  [A_{\mu\lambda},\eta_{\nu}]
-i g [A_{\mu\nu\lambda},\eta])\nonumber\\
&=&-ig\{~\partial_{\mu} ( [\eta,\xi_{\nu\lambda}] + [\eta_{\nu},\xi_{\lambda}]
+ [\eta_{\lambda},\xi_{\nu}] + [\eta_{\nu\lambda},\xi])\nonumber\\
&~&~-ig [A_{\mu},([\eta,\xi_{\nu\lambda}]  +
[\eta_{\nu},\xi_{\lambda}]+ [\eta_{\lambda},\xi_{\nu}]+
[\eta_{\nu\lambda},\xi])]\nonumber\\
&~&~-i g  [A_{\mu\nu},( [\eta,\xi_{\lambda}]  + [\eta_{\lambda},\xi])]
-i g  [A_{\mu\lambda},( [\eta,\xi_{\nu}]  + [\eta_{\nu},\xi])]
-i g[A_{\mu\nu\lambda},[\eta,\xi] ] \}\nonumber\\
&=&-ig~\{~\partial_{\mu} \zeta_{\nu\lambda} -i g[A_{\mu},\zeta_{\nu\lambda}]
-i g  [A_{\mu\nu},\zeta_{\lambda}]-i g   [A_{\mu\lambda},\zeta_{\nu}]-
i g [A_{\mu\nu\lambda},\zeta]~\} =\delta_{\zeta}A_{\mu\nu\lambda},\nonumber
\eeqa
where
\beqa\label{commutatorofparameterslow}
\zeta =[\eta,\xi],~~~\zeta_{\nu} = [\eta,\xi_{\nu}] +[\eta_{\nu},\xi],~~~~
\zeta_{\nu\lambda} = [\eta,\xi_{\nu\lambda}] +  [\eta_{\nu},\xi_{\lambda}]
+ [\eta_{\lambda},\xi_{\nu}]+[\eta_{\nu\lambda},\xi].
\eeqa
The fact that our extended gauge transformations form a closed algebra is essentially based
on a specific set of tensor fields and on the fact that the tensor $A_{\mu\nu}$
does not have any restrictions on its symmetry properties and therefore
belongs to a reducible representation of the Poincar\'e algebra. As it is well known,
this phenomena also takes place in the case of Dirac equation, where the wave function
belongs to a reducible representation.

The invariance of the Lagrangian ${{\cal L}}_2$
can be proved by realizing that the extended gauge transformations
induce  the transformation of the new field strengths of the
form
\beqa\label{fieldstrengh3thtransfor}
\delta G^{a}_{\mu\nu,\lambda} &=& g f^{abc} (~G^{b}_{\mu\nu,\lambda} \xi^c
+ G^{b}_{\mu\nu} \xi^{c}_{\lambda}~),\nonumber\\
\delta G^{a}_{\mu\nu,\lambda\rho} &=& g f^{abc}
(~G^{b}_{\mu\nu,\lambda\rho} \xi^c
+ G^{b}_{\mu\nu,\lambda} \xi^{c}_{\rho} +
G^{b}_{\mu\nu,\rho} \xi^{c}_{\lambda} +
G^{b}_{\mu\nu} \xi^{c}_{\lambda\rho}~).
\eeqa
One can establish this transformation law for the
field strengths $G^{a}_{\mu\nu,\lambda}$ and $G^{a}_{\mu\nu,\lambda\rho}$
by direct calculation using the definition of the field strengths
(\ref{fieldstrength2}), (\ref{fieldstrength3}), the transformation laws
for the tensor gauge fields (\ref{polygauge}),(\ref{matrixform})
and the Jaboci identity,
which holds for matrices.  Now we can prove the
invariance of the Lagrangian ${{\cal L}}_2$. Indeed, its variation
is
\beqa
\delta {{\cal L}}_2 =
&-&{1\over 2} G^{a}_{\mu\nu,\lambda} g f^{abc} (G^{b}_{\mu\nu,\lambda} \xi^c  +
G^{b}_{\mu\nu} \xi^{c}_{\lambda})
-{1\over 4} g f^{abc} G^{b}_{\mu\nu} \xi^c G^{a}_{\mu\nu,\lambda\lambda}\nonumber\\
&-&{1\over 4} G^{a}_{\mu\nu} g f^{abc} (G^{b}_{\mu\nu,\lambda\lambda} \xi^c  +
2 G^{b}_{\mu\nu, \lambda} \xi^{c}_{\lambda}+
G^{b}_{\mu\nu} \xi^{c}_{\lambda \lambda})=0 .\nonumber
\eeqa
The two terms in the above expression are
equal to zero because they are symmetric in Lorentz indices
and are antisymmetric in group indices, like the terms
$G^{a}_{\mu\nu,\lambda} g f^{abc} G^{b}_{\mu\nu,\lambda} \xi^c$
and $G^{a}_{\mu\nu} g f^{abc}G^{b}_{\mu\nu} \xi^{c}_{\lambda \lambda}$.
The other four terms show delicate cancellation, like terms of the form
$G^{a}_{\mu\nu,\lambda} g f^{abc} G^{b}_{\mu\nu} \xi^{c}_{\lambda}$
and $g f^{abc} G^{b}_{\mu\nu} \xi^c G^{a}_{\mu\nu,\lambda\lambda}$. This follows
from the antisymmetric nature of the group structure constants
$f^{abc}$.

In order to describe the interaction of the Yang-Mills boson and
the rank-2 gauge boson system one should
add the Yang-Mills Lagrangian ${{\cal L}}_1$  to ${{\cal L}}_2$. This yields
\be\label{actiontwo}
{{\cal L}} =  {{\cal L}}_1 +  g_2 {{\cal L}}_2 =-{1\over 4}G^{a}_{\mu\nu}
G^{a}_{\mu\nu} + g_2 \{
-{1\over 4}G^{a}_{\mu\nu,\lambda}G^{a}_{\mu\nu,\lambda}
-{1\over 4}G^{a}_{\mu\nu}G^{a}_{\mu\nu,\lambda\lambda}\}.
\ee
Both terms are invariant with respect to the extended gauge transformations
(\ref{polygauge}).
It is important that i) the Lagrangian does not
contain higher derivatives of higher-rank gauge fields, as it was suggested in different
approaches to the higher-spin field theories, and that
ii) all interactions take place
through three- and four-particle exchanges iii) the extended gauge transformations
do not define the coupling constants $g_{2}$.
The coupling constant $g_{2}$  remains arbitrary because
every term in the sum is separately gauge invariant and the extended gauge symmetry
alone does not define it.

The immediate conclusion which
can be drawn from above properties is that the extended gauge theory will have
the same degree of divergence of its Feynman diagrams as the Yang-Mills theory
does and most probably will be renormalizable.  The other conclusion is that
the second-rank tensor gauge field $A^{a}_{\mu\nu}$, which in our theory
is an arbitrary
nonsymmetric tensor $A^{a}_{\mu\nu} \neq A^{a}_{\nu\mu}$,
does not directly coincide with the graviton,
because it is a charged gauge field and it has different
gauge symmetries and interactions.
We shall discuss this question also in the tenth section.

Let us now consider the equations which follow from this Lagrangian. The equations of
motion that follow for the Yang-Mills fields will be modified by additional terms
which we can obtain by variation of the full action (\ref{actiontwo}) with respect to the
gauge field $A^{a}_{\mu}$:
$$
\nabla^{ab}_{\mu}G^{b}_{\mu\nu} + g_2 (g f^{acb} A^{c}_{\mu\lambda} G^{b}_{\mu\nu,\lambda} +
{1\over 2} g f^{acb}A^{c}_{\mu\lambda\lambda} G^{b}_{\mu\nu} +
{1\over 2 }\nabla^{ab}_{\mu} G^{b}_{\mu\nu,\lambda\lambda})=J^{a}_{\nu}.
$$
The first term represents the classical Yang-Mills operator, whereas the rest of the
terms represent the modification of the Yang-Mills equations by the higher-rank gauge fields.
One should also derive the new field equations varying the action
with respect to the second-rank gauge field $A^{a}_{\mu\nu}$
$$
\nabla^{ab}_{\mu}G^{b}_{\mu\nu,\lambda} +
g f^{acb} A^{c}_{\mu\lambda} G^{b}_{\mu\nu} =J^{a}_{\nu\lambda}.
$$
Remarkably enough, the variation of the action with respect to the third-rank gauge field
$A^{a}_{\mu\nu\lambda}$ will give the Yang-Mills equations
$$
\nabla^{ab}_{\mu}G^{b}_{\mu\nu}  =J^{a}_{\nu\lambda\lambda}.
$$

The Lagrangian ${{\cal L}}_1 +  g_2 {{\cal L}}_2$ contains the
third-rank gauge fields $A^{a}_{\mu\lambda\lambda}$, but without corresponding
kinetic terms, therefore they play mostly the role of the auxiliary fields.
In order to make the fields $A^{a}_{\mu\nu\lambda}$ dynamical,
we shall proceed introducing the corresponding
kinetic term, which should be quadratic in the field derivatives. With that aim
we shall further generalize the gauge transformations
for the higher-rank gauge fields in the next section.

\section{\it Third-Rank Tensor Gauge Fields}

In order to describe dynamics of the third-rank gauge field
$A^{a}_{\mu\nu\lambda}$ we shall
introduce additional tensor gauge fields $A^{a}_{\mu\nu\lambda\rho}$ and
$A^{a}_{\mu\nu\lambda\rho\sigma}$. These tensors are totally symmetric
with respect to the indices $\nu\lambda\rho$ and
$\nu\lambda\rho\sigma$ respectively, but have no symmetries with
respect to the permutations of the first index $\mu$.
Using these gauge fields we shall define the action as:
\beqa
{{\cal L}}_3 =-{1\over 4}G^{a}_{\mu\nu,\lambda\rho}G^{a}_{\mu\nu,\lambda\rho}
-{1\over 8}G^{a}_{\mu\nu ,\lambda\lambda}G^{a}_{\mu\nu ,\rho\rho}
-{1\over 2}G^{a}_{\mu\nu,\lambda}  G^{a}_{\mu\nu ,\lambda \rho\rho}
-{1\over 8}G^{a}_{\mu\nu}  G^{a}_{\mu\nu ,\lambda \lambda\rho\rho}~,
\eeqa
where the new field strengths for the tensor gauge fields are
\beqa\label{spin4fieldstrenghth}
G^{a}_{\mu\nu ,\lambda \rho \sigma} =
\partial_{\mu} A^{a}_{\nu \lambda \rho \sigma} -
\partial_{\nu} A^{a}_{\mu \lambda\rho\sigma} +
g f^{abc}\{~A^{b}_{\mu}~A^{c}_{\nu \lambda \rho\sigma}
+A^{b}_{\mu\lambda}~A^{c}_{\nu\rho \sigma} +
A^{b}_{\mu\rho }~A^{c}_{\nu\lambda\sigma} +
A^{b}_{\mu\sigma}~A^{c}_{\nu\lambda\rho} +\nn\\
+A^{b}_{\mu\lambda\rho}~A^{c}_{\nu \sigma} +
A^{b}_{\mu\lambda\sigma}~A^{c}_{\nu\rho} +
A^{b}_{\mu\rho\sigma}~A^{c}_{\nu \lambda} +
     A^{b}_{\mu\lambda\rho\sigma }~A^{c}_{\nu} ~\}\nonumber
\eeqa
and
\beqa\label{spin4fieldstrenghth4}
G^{a}_{\mu\nu ,\lambda \rho \sigma\delta} =
\partial_{\mu} A^{a}_{\nu \lambda \rho \sigma\delta} -
\partial_{\nu} A^{a}_{\mu \lambda\rho\sigma\delta} +
g f^{abc}\{~A^{b}_{\mu}~A^{c}_{\nu \lambda \rho\sigma\delta}
+\sum_{ \lambda \leftrightarrow \rho,\sigma,\delta}
        A^{b}_{\mu\lambda}~A^{c}_{\nu\rho \sigma\delta} +~~~~~~~~~~~~~~~~~~~~~~\\
+\sum_{\lambda,\rho \leftrightarrow \sigma,\delta}
        A^{b}_{\mu\lambda\rho}~A^{c}_{\nu\sigma\delta} +
\sum_{\lambda,\rho,\sigma\leftrightarrow \delta}
       A^{b}_{\mu\lambda\rho\sigma}~A^{c}_{\nu\delta} +
     A^{b}_{\mu\lambda\rho\sigma\delta }~A^{c}_{\nu} ~\},\nonumber
\eeqa
where the terms in parenthesis have been symmetrized over $\lambda \rho\sigma$ and
$\lambda \rho \sigma\delta$ respectively. The extended field strength tensors are
antisymmetric in their first two indices $\mu,\nu$ and are totally
symmetric with respect to the rest of the indices:
$$
G^{a}_{\mu\nu,\lambda\rho\sigma....} = -G^{a}_{\nu\mu,\lambda\rho\sigma....},~~~
G^{a}_{\mu\nu,\lambda\rho\sigma....} = G^{a}_{\mu\nu,\rho\lambda\sigma....}=.....
$$

We shall prove that the Lagrangian ${{\cal L}}_3$ is invariant under
the extended gauge transformation which we shall define for the fourth-rank
gauge field as
\beqa\label{gaugetransform4}
\delta_{\xi}  A_{\mu\nu\lambda\rho} =\partial_{\mu}\xi_{\nu\lambda\rho}
-i g[A_{\mu},\xi_{\nu\lambda\rho}]
-i g [A_{\mu\nu},\xi_{\lambda\rho}]
-i g [A_{\mu\lambda},\xi_{\nu\rho}]
-i g [A_{\mu\rho},\xi_{\nu\lambda}]-\\
-i g  [A_{\mu\nu\lambda},\xi_{\rho}]
-i g  [A_{\mu\nu\rho},\xi_{\lambda}]
-i g  [A_{\mu\lambda\rho},\xi_{\nu}]
-i g [A_{\mu\nu\lambda\rho},\xi]\nonumber
\eeqa
and for the fifth-rank tensor gauge field as
\beqa\label{gaugetransform5}
\delta_{\xi}  A_{\mu\nu\lambda\rho\sigma} &=&\partial_{\mu}\xi_{\nu\lambda\rho\sigma}
-i g[A_{\mu},\xi_{\nu\lambda\rho\sigma}]
-i g \sum_{\nu \leftrightarrow \lambda\rho\sigma} [A_{\mu\nu},\xi_{\lambda\rho\sigma}]-\\
   &~&-ig\sum_{\nu\lambda \leftrightarrow \rho\sigma} [A_{\mu\nu\lambda},\xi_{\rho\sigma}]
      -ig\sum_{\nu\lambda\rho \leftrightarrow \sigma} [A_{\mu\nu\lambda\rho},\xi_{\sigma}]
-i g [A_{\mu\nu\lambda\rho},\xi], ~\nonumber
\eeqa
where the gauge parameters $\xi_{\nu\lambda\rho}$ and $\xi_{\nu\lambda\rho\sigma}$
are totally symmetric rank-3 and rank-4 tensors. These gauge transformations preserve the
symmetries of the gauge fields, because the r.h.s are symmetric with respect to
the indices $\nu\lambda\rho$ and $\nu\lambda\rho\sigma$.
One should be convinced that this transformation again forms a closed algebra. The
commutator of two gauge transformations acting on a rank-4 gauge field is:
\beqa
[\delta_{\eta},\delta_{\xi}]A_{\mu\nu\lambda\rho} &=&
\delta_{\eta}~(-i g[A_{\mu},\xi_{\nu\lambda\rho}]
-i g  \sum_{\nu \leftrightarrow \lambda\rho}[A_{\mu\nu},\xi_{\lambda\rho}]
    -i g \sum_{\nu\lambda \leftrightarrow \rho} [A_{\mu\nu\lambda},\xi_{\rho}]-
          i g [A_{\mu\nu\lambda\rho},\xi]) \nonumber\\
&-& \delta_{\xi}~(-i g[A_{\mu},\eta_{\nu\lambda\rho}]
-i g \sum_{\nu \leftrightarrow \lambda\rho} [A_{\mu\nu},\eta_{\lambda\rho}]
     -i g  \sum_{\nu\lambda \leftrightarrow \rho}[A_{\mu\nu\lambda},\eta_{\rho}]
         -i g [A_{\mu\nu\lambda\rho},\eta])\nonumber\\
= -ig~\{~\partial_{\mu} \zeta_{\nu\lambda\rho} &-&i g[A_{\mu},\zeta_{\nu\lambda\rho}]
        -i g \sum_{\nu \leftrightarrow \lambda\rho} [A_{\mu\nu},\zeta_{\lambda\rho}]
        -i g \sum_{\nu\lambda \leftrightarrow \rho}  [A_{\mu\nu\lambda},\zeta_{\rho}]-
              i g [A_{\mu\nu\lambda\rho},\zeta]~\}\nonumber
\eeqa
and is equivalent to the extended gauge transformation with the gauge parameters of the form
\beqa\label{commutatorofparameters}
\zeta =[\eta,\xi],~~~\zeta_{\nu} = [\eta,\xi_{\nu}] +[\eta_{\nu},\xi],~~~~
\zeta_{\nu\lambda} = [\eta,\xi_{\nu\lambda}] +  [\eta_{\nu},\xi_{\lambda}]
+ [\eta_{\lambda},\xi_{\nu}]+[\eta_{\nu\lambda},\xi],\nonumber\\
\\
\zeta_{\nu\lambda\rho} = [\eta,\xi_{\nu\lambda\rho}] +
\sum_{\nu \leftrightarrow \lambda\rho} [\eta_{\nu},\xi_{\lambda\rho}]
+\sum_{\nu\lambda \leftrightarrow \rho}~[\eta_{\nu\lambda},\xi_{\rho}]+
  [\eta_{\nu\lambda\rho},\xi ]~.~~~~~~~~~~~~~~~~~\nonumber
\eeqa
The commutator of two gauge transformations acting on rank-5 gauge field
is again equivalent to a gauge transformation with the gauge parameters:
\beqa\label{gaugetransform6}
\zeta_{\nu\lambda\rho\sigma} = [\eta,\xi_{\nu\lambda\rho\sigma}] +
      \sum_{\nu \leftrightarrow \lambda\leftrightarrow\rho\leftrightarrow\sigma}
      (~[\eta_{\nu},\xi_{\lambda\rho\sigma}]
          +[\eta_{\nu\lambda},\xi_{\rho\sigma}]
             +[\eta_{\nu\lambda\rho},\xi_{\sigma}]~)~+~
  [\eta_{\nu\lambda\rho\sigma},\xi].
\eeqa
One can express all commutators of extended gauge transformations acting on
a tensor gauge fields in the form
\be
[\delta_{\eta},\delta_{\xi}]A_{\mu\nu\lambda\rho\sigma} = -i g~
 \delta_{\zeta} A_{\mu\nu\lambda\rho\sigma}~,
\ee
where $\zeta$ is given by the expressions (\ref{commutatorofparameters})
and (\ref{gaugetransform6}).

The extended gauge transformation of the higher rank tensor gauge fields (\ref{gaugetransform4})
and (\ref{gaugetransform5}) induces
the gauge transformation of the fields strengths of the form
\beqa\label{spin4fieldstrenghthtransfor}
\delta G^{a}_{\mu\nu,\lambda\rho\sigma} =
g f^{abc} (~G^{b}_{\mu\nu,\lambda\rho\sigma} ~\xi^c  +
 G^{b}_{\mu\nu, \lambda\rho} ~\xi^{c}_{\sigma}+
G^{b}_{\mu\nu, \lambda\sigma} ~\xi^{c}_{\rho}+
G^{b}_{\mu\nu, \rho\sigma} ~\xi^{c}_{\lambda}+~~~~~~~~~~~~~~~~~~~~~~~~~\\
+ G^{b}_{\mu\nu,\lambda } ~\xi^{c}_{\rho\sigma}+
 G^{b}_{\mu\nu,\rho} ~\xi^{c}_{\lambda\sigma}+
 G^{b}_{\mu\nu,\sigma} ~\xi^{c}_{\lambda \rho}+
G^{b}_{\mu\nu } ~\xi^{c}_{\lambda\rho\sigma}~)\nonumber
\eeqa
and
\beqa\label{fieldstrengh5thtransfor}
\delta G^{a}_{\mu\nu,\lambda\rho\sigma\delta} =
g f^{abc} (~G^{b}_{\mu\nu,\lambda\rho\sigma\delta} ~\xi^c
+ \sum_{ \lambda\rho,\sigma \leftrightarrow \delta}
G^{b}_{\mu\nu, \lambda\rho\sigma} ~\xi^{c}_{\delta}
+~~~~~~~~~~~~~~~~~~~~~~~~~~~~~~~~~~~~~~~~~~~~~\\
+ \sum_{ \lambda\rho \leftrightarrow \sigma,\delta} G^{b}_{\mu\nu, \lambda\rho} ~\xi^{c}_{\sigma\delta}+
         \sum_{ \lambda \leftrightarrow \rho,\sigma,\delta}
         G^{b}_{\mu\nu, \lambda} ~\xi^{c}_{\rho\sigma\delta}~+
         G^{b}_{\mu\nu } ~\xi^{c}_{\lambda\rho\sigma\delta}).\nonumber
\eeqa
Using the above homogeneous transformations for the field strengths
it is easy to demonstrate the invariance of the
Lagrangian ${{\cal L}}_3$ with respect to the extended gauge transformations
in full analogy with the calculations of the previous section.

In the Lagrangian ${{\cal L}}_3$
the third-rank tensor gauge field is dynamical and the fields $A^{a}_{\mu} $,
$A^{a}_{\mu\nu} $, $A^{a}_{\mu\nu\lambda\rho} $ and
$A^{a}_{\mu\nu\lambda\rho\omega}$ are not. To make the lower-rank gauge
fields dynamical one should add the corresponding low-order Lagrangians, thus
\be
{{\cal L}} =  {{\cal L}}_1 +  g_2 {{\cal L}}_2  + g_3 {{\cal L}}_3,
\ee
where $g_2$ and $g_3$ are new coupling constants.
The tensor gauge fields $A_{\mu\nu\lambda\rho}$
and $A_{\mu\nu\lambda\rho\sigma}$ remain auxiliary.

The last two sections provide a clear hint how to proceed with the extension
of the above construction. Therefore our intention in the next section will be
to give a general formulation of the extended gauge transformation for the arbitrary
tensor gauge fields, to define the corresponding field strengths and the invariant
Lagrangian. This will complete the formulation of the extended
gauge principle for arbitrary tensor gauge fields.

\section{{\it Non-Abelian Tensor Gauge Fields }}

In a general case we shall
consider  tensor gauge fields
$A^{a}_{\mu\lambda_1 ... \lambda_{s}}$ of the rank
$s+1$, which are totally symmetric with respect to the
indices $  \lambda_1 ... \lambda_{s}  $, but a priory have no symmetries with
respect to the first index  $\mu$.  The corresponding
field strength we shall define by the following expression:
\be\label{fieldstrengthgeneral}
G^{a}_{\mu\nu ,\lambda_1 ... \lambda_{s}} =
\partial_{\mu} A^{a}_{\nu \lambda_1 ... \lambda_{s}} -
\partial_{\nu} A^{a}_{\mu \lambda_1 ... \lambda_s} +
g f^{abc}\sum^{s}_{i=0}~ \sum_{P's} ~A^{b}_{\mu \lambda_1 ... \lambda_i}~
A^{c}_{\nu \lambda_{i+1} ... \lambda_{s}},
\ee
where the sum $\sum_{P's}$ runs over all permutations of two
sets of indices $\lambda_1 ... \lambda_i$ and $\lambda_{i+1} ... \lambda_{s}$
which correspond to nonequal terms.
All permutations of indices within two sets $\lambda_1 ... \lambda_i$ and
$\lambda_{i+1} ... \lambda_{s}$ correspond to equal terms, because
gauge fields are totally symmetric with respect to $\lambda_1 ... \lambda_i$ and
$\lambda_{i+1} ... \lambda_{s}$. Therefore there are
$$
\frac{s!}{i!(s-i)!}
$$
nonequal terms in the sum $\sum_{P's}$. Thus in the sum $\sum_{P's}$ there is one
term of the form $A_{\mu}A_{\nu\lambda_1\lambda_{2} ... \lambda_{s}}$, there are
$s$ terms
of the form $A_{\mu \lambda_1}
A_{\nu \lambda_{2} ... \lambda_{s}}$ and $s(s-1)/2$ terms of the form
$A_{\mu \lambda_1 \lambda_2}~
A_{\nu \lambda_{3} ... \lambda_{s}}$ and so on.

In the above definition of the extended gauge field strength
$G^{a}_{\mu\nu ,\lambda_1 ... \lambda_{s}}$,
together with the classical Yang-Mills gauge boson
$A^{a}_{\mu}$, there participate the set of higher-rank gauge fields
$A^{a}_{\mu\lambda_1}$, $A^{a}_{\mu\lambda_1 , \lambda_2}$,  ... ,
$A^{a}_{\mu\lambda_1 ...\lambda_{s}}$ up to the rank $s+1$.
By construction the field strength (\ref{fieldstrengthgeneral})
is antisymmetric with respect to its first two indices
\be\label{anisymmetry}
G^{a}_{\mu\nu ,\lambda_1 ... \lambda_{s}}~ = ~-~G^{a}_{\nu \mu,\lambda_1 ... \lambda_{s}}
\ee
and is totally symmetric with respect to the rest of the indices
$\lambda_1 ... \lambda_{s}$
\be\label{totalsymmetry}
G^{a}_{\mu\nu ,..\lambda_{i}...\lambda_{j}.. } =
G^{a}_{\mu\nu ,..\lambda_{j}...\lambda_{i}..}~~~~~~~~,
\ee
where $1 \leq i < j \leq s$.
The extended gauge transformation of the fields we shall define by the
formula
\be\label{generalgaugetransform}
\delta A^{a}_{\mu\lambda_1 ... \lambda_s} = ( \delta^{ab}\partial_{\mu}
+g f^{acb} A^{c}_{\mu})\xi^{b}_{\lambda_1\lambda_2 ...\lambda_s} +
g f^{acb}~\sum^{s}_{i=1}  \sum_{P's} A^{c}_{\mu\lambda_1 ...\lambda_i}
\xi^{b}_{\lambda_{i+1} ...\lambda_s },
\ee
where the infinitesimal gauge parameters $\xi^{b}_{\lambda_{1} ...\lambda_s }$ are
totally symmetric rank-s tensors. The summation  $\sum_{P's}$ is again
over all permutations of two
sets of indices $\lambda_1 ... \lambda_i$ and $\lambda_{i+1} ... \lambda_{s}$
which correspond to nonequal terms. It is obvious that this transformation
preserves the symmetry properties of the tensor gauge field
$A^{a}_{\mu\lambda_1 ... \lambda_s}$. Indeed, the first term in the r.h.s. is a covariant
derivative of the totally symmetric rank-s tensor
$\nabla^{ab}_{\mu}\xi^{b}_{\lambda_1\lambda_2 ...\lambda_s}$ and every term
$\sum_{P's} A^{c}_{\mu\lambda_1 ...\lambda_i}
\xi^{b}_{\lambda_{i+1} ...\lambda_s }$ in the second sum is totally
symmetric with respect to the indices $\lambda_1\lambda_2 ...\lambda_s$ by construction.
The matrix form of the transformation has the form:
\beqa
\delta_{\xi} A_{\mu\lambda_1 ... \lambda_s} &=& \partial_{\mu}
\xi_{\lambda_1\lambda_2 ...\lambda_s}
-i g  [A_{\mu}, \xi_{\lambda_1\lambda_2 ...\lambda_s}] -i g
\sum^{s}_{i=1}  \sum_{P's} [A_{\mu\lambda_1 ...\lambda_i},
\xi_{\lambda_{i+1} ...\lambda_s }].
\eeqa

As we demonstrated in the cases of lower-rank gauge fields, the extended gauge
transformations form a closed algebra. This remains true in the general case as well.
Indeed, the commutator of two extended  infinitesimal gauge transformations
can be expressed in the form
\be
[~\delta_{\eta}~,~\delta_{\xi}]~A_{\mu\lambda_1\lambda_2 ...\lambda_s} ~=~
-i g~ \delta_{\zeta} ~A_{\mu\lambda_1\lambda_2 ...\lambda_s}~,
\ee
where the gauge parameters $\{\zeta\}$ are given by the symmetric commutators of
two gauge parameters of the form (\ref{commutatorofparameters}),(\ref{gaugetransform6}):
\beqa\label{generalalgebralow}
\zeta_{\lambda_1\lambda_2 ... \lambda_i} =
\sum^{i}_{j=0}  \sum_{P's} [~\eta_{\lambda_1 ...\lambda_j},
\xi_{\lambda_{j+1} ...\lambda_i }~]~~~~~~~~~,i=0,1,...s
\eeqa
and the sum $\sum_{P's}$ runs over all nonequal permutations of two
sets of indices  $\lambda_1 ... \lambda_j$ and $\lambda_{j+1} ... \lambda_{i}$
(see Appendix).

The inhomogeneous extended gauge transformation (\ref{generalgaugetransform})
induces the homogeneous gauge
transformation of the corresponding field strength (\ref{fieldstrengthgeneral}) of the form
\be
\delta G^{a}_{\mu\nu,\lambda_1 ... \lambda_s} =
g f^{abc} \sum^{s}_{i=0}~ \sum_{P's} ~G^{b}_{\mu\nu,\lambda_1 ... \lambda_i}
\xi^{c}_{\lambda_{i+1}...\lambda_s}.
\ee
The symmetry properties (\ref{anisymmetry}) and
(\ref{totalsymmetry}) of the field strength  $G^{a}_{\mu\nu,\lambda_1 ... \lambda_s}$
remain invariant in the course of this transformation.

The polygauge invariant Lagrangian  now can be formulated in the form
\beqa\label{fulllagrangian1}
{{\cal L}}_{s+1}&=&-{1\over 4} ~
G^{a}_{\mu\nu, \lambda_1 ... \lambda_s}~
G^{a}_{\mu\nu, \lambda_{1}...\lambda_{s}} +.......\nonumber\\
&=& -{1\over 4}\sum^{2s}_{i=0}~a^{s}_i ~
G^{a}_{\mu\nu, \lambda_1 ... \lambda_i}~
G^{a}_{\mu\nu, \lambda_{i+1}...\lambda_{2s}}
\sum_{p's} \eta^{\lambda_{i_1} \lambda_{i_2}} .......
\eta^{\lambda_{i_{2s-1}} \lambda_{i_{2s}}}~,
\eeqa
where the sum $\sum_p$ runs over all permutations of indices
$\lambda_1 ... \lambda_{2s}$ which correspond to nonequal terms. For the low
values of $s=0,1,2,...$ the numerical coefficients $$a^{s}_i = {s!\over i!(2s-i)!}$$
are: $a^{0}_0=1;~~a^{1}_1 =1,a^{1}_0 =a^{1}_2 =1/2;~~
a^{2}_2 =1/2,a^{2}_1 =a^{2}_3 =1/3,a^{2}_0 =a^{2}_4 =1/12;$ and so on.
In order to describe fixed rank-$(s+1)$ gauge field
one should have  at disposal all gauge fields
up to the rank $2s+1$.
In order to make all tensor gauge fields dynamical one should add
the corresponding kinetic terms. Thus the invariant
Lagrangian describing dynamical tensor gauge bosons of all ranks
has the form
\be\label{fulllagrangian2}
{{\cal L}} = \sum^{\infty}_{s=1}~ g_s {{\cal L}}_s~,
\ee
where $g_s$~$( s=1,2,...$) are new coupling constants.

As in the case of the Yang-Mills theory \cite{yang},
the Feynman diagrams here have three elementary types of vertices.
The "primitive divergences" are also in a finite number, because the superficial degree
of divergence $d$ for Feynman diagrams with $N_B$ bosonic and $N_F$ fermionic
external lines is $d=-{D-2\over 2}N_B - {D-1\over 2}N_F + D$.
The index of divergence $r= {D-2\over 2}b +{D-1\over 2}f + \delta -D $ of the
interaction Lagrangian ${{\cal L}}_{int} \sim g (\partial)^\delta (\phi)^b (\psi)^f$
is equal to zero for all vertices. Here D is dimension of the space-time, b - number of
boson fields, f - number of fermion fields and $\delta$ is the
number of space-time derivatives in ${{\cal L}}_{int}$. The conclusion which
can be drawn from this property is that the extended gauge theory has
the same degree of divergence of its Feynman diagrams as the Yang-Mills theory
does and most probably will be renormalizable in its general form.

\section{{\it Interaction of Fermions }}

In this section we shall introduce gauge invariant interaction of the
tensor gauge fields with higher-rank fermion fields representing
matter. We shall use the Rarita-Schwinger spin-tensor fields
$\psi^{\alpha}_{\lambda_{1},...,\lambda_{s}}$, where $\alpha$ is
a spinor index and $\lambda_{1},...,\lambda_{s}$ are vector indices. Tensors
are totally symmetric with respect to the indices $\lambda_{1},...,\lambda_{s}$.
The transformation of the fermions under the extended isotopic group we
shall define by the formulas \cite{Savvidy:2003fx}
\beqa\label{mattertransformation}
\delta_{\xi}  \psi  &=&-i \sigma^{a}\xi^{a}  \psi ,\nonumber\\
\delta_{\xi}  \psi_{\lambda} &=& -i \sigma^{a}( \xi^{a} ~\psi_{\lambda}   +
\xi^{a}_{\lambda}~ \psi) ,\nonumber\\
\delta_{\xi}  \psi_{\lambda\rho} &=& -i \sigma^{a}( \xi^{a} ~\psi_{\lambda\rho}   +
\xi^{a}_{\lambda}~ \psi_{\rho} + \xi^{a}_{\rho}~
\psi_{\lambda} + \xi_{\lambda\rho} ~ \psi),\\
........&.&.......................,\nn
\eeqa
where $\sigma^{a}$ are the matrices of the
representation $\sigma$ of the semisimple Lie
group G, according to which all $\psi's$ are transforming. The general form of the
above transformation is:
\beqa
\delta_{\xi}  \psi_{\lambda_1 ... \lambda_{s}}(x) &=& -i \sum^{s}_{i=0}  \sum_{P's}
 \xi_{\lambda_1 ... \lambda_{i}} ~\psi_{\lambda_{i+1} ... \lambda_{s}}(x) ,~~~~~~~
 s=0,1,2,...,
\eeqa
where we are using the same notations as in the previous section.
The transformation properties
of the covariant derivatives are:
\beqa\label{covariantmattertransformation}
\delta_{\xi}  \nabla_{\mu} \psi  &=&-i  \xi  _{\mu} \nabla\psi ,\nonumber\\
\delta_{\xi}  \nabla_{\mu} \psi_{\lambda} &=& -i \xi \nabla_{\mu} \psi_{\lambda}   -i
\xi_{\lambda} \nabla_{\mu} \psi -i \nabla_{\mu} \xi_{\lambda} ~\psi ,\nonumber\\
\delta_{\xi}  \nabla_{\mu}\psi_{\lambda\rho} &=& -i  \xi\nabla_{\mu}\psi_{\lambda\rho}  -i
\xi_{\lambda}~ \nabla_{\mu}\psi_{\rho} -i \xi_{\rho}\nabla_{\mu}\psi_{\lambda}-i
\xi_{\lambda\rho}\nabla_{\mu}~ \psi \nn \\ &~&-i
\nabla_{\mu}\xi_{\lambda}~ \psi_{\rho} -i
\nabla_{\mu}\xi_{\rho}~ \psi_{\lambda} -i
\nabla_{\mu}\xi_{\lambda\rho}~ \psi,...\\
...........&.&.......................,\nn
\eeqa
where we are using the matrix notation for the gauge fields
$A_{\mu} = \sigma^{a} A^{a}_{\mu}$ and gauge parameters $\xi=\sigma^{a}\xi^{a}$.
In this section we shall redefine the gauge parameters
$\xi \rightarrow -{1\over g} ~\xi$, thus  $U(\xi) \rightarrow exp(- i \sigma^{a}\xi^{a} )$ and
\beqa\label{matrixformredefined}
\delta_{\xi}  A_{\mu} &=& -{1\over g}(\partial_{\mu}\xi -i g[A_{\mu},\xi]),\nonumber\\
\delta_{\xi}  A_{\mu\nu} &=&-{1\over g}( \partial_{\mu}\xi_{\nu} -i g[A_{\mu},\xi_{\nu}]
-i g [A_{\mu\nu},\xi])
\eeqa
and so on.

The invariant Lagrangian for spin-1/2 matter is
\be
{{\cal L}}_{1/2} =   \bar{\psi}  \gamma_{\mu} ( i\partial_{\mu} ~+~
g \sigma^{a} A^{a}_{\mu} )\psi = \bar{\psi}  ( i \not\!\partial +
g \not\!\!A )\psi.
\ee
To describe the dynamics of the spin-vector field $\psi_{\mu}$
we shall introduce additional
rank-2 spin-tensor $\psi_{\mu\nu}$. The invariant Lagrangian has the form:
\beqa\label{fermionlagrangian}
{{\cal L}}_{3/2} &=&
\bar{\psi}_{\lambda} \gamma_{\mu} ( i\partial_{\mu} + g A_{\mu} )\psi_{\lambda} +
{1\over 2}\bar{\psi} \gamma_{\mu} (i\partial_{\mu} + g A_{\mu} )\psi_{\lambda\lambda}+
{1\over 2}\bar{\psi}_{\lambda\lambda} \gamma_{\mu}
(i\partial_{\mu} + g A_{\mu} )\psi\nonumber\\
&+& g \bar{\psi}_{\lambda} \gamma_{\mu}  A_{\mu\lambda} \psi
+g \bar{\psi}  \gamma_{\mu} A_{\mu\lambda} \psi_{\lambda}
+{1\over 2}g \bar{\psi} \gamma_{\mu}  A_{\mu\lambda\lambda} \psi ~.
\eeqa
We have to prove that the Lagrangian is invariant under simultaneous
extended gauge transformations of the fermions and tensor gauge fields.
It is easier to use compact notations in order to simplify algebraic computations:
\beqa
{{\cal L}}_{3/2} &=&
\bar{\psi}_{\lambda}   ( i\not\!\partial + g\not\!\!A  )\psi_{\lambda} +
{1\over 2}\bar{\psi}  (i\not\!\partial + g \not\!\!A )\psi_{\lambda\lambda}+
{1\over 2}\bar{\psi}_{\lambda\lambda} (i\not\!\partial + g \not\!\!A )\psi\nonumber\\
&+& g \bar{\psi}_{\lambda}   \not\!\!A_{\lambda} \psi
+g \bar{\psi}  \not\!\!A_{\lambda} \psi_{\lambda}
+{1\over 2}~g \bar{\psi}    \not\!\!A_{\lambda\lambda} \psi,
\eeqa
where $\not\!\partial= \gamma^{\mu}  \partial_{\mu}$,~
$\not\!\!A = \gamma^{\mu}  A_{\mu} $,~
$\not\!\!A_{\lambda} = \gamma^{\mu}  A_{\mu\lambda} $ and
$\not\!\!A_{\lambda\lambda} = \gamma^{\mu}  A_{\mu\lambda\lambda} $.

The variation of the first three terms of the Lagrangian  results in the
expression
$$
\delta {{\cal L}}_{3/2}(I+II+III) ={1\over 2}\bar{\psi}
(\!\not\!\partial \xi_{\lambda\lambda}
-i g [\not\!\!A,\xi_{\lambda\lambda}])\psi +
\bar{\psi}(\!\not\!\partial \xi_{\lambda}
-i g [\not\!\!A,\xi_{\lambda}])\psi_{\lambda} +
\bar{\psi}_{\lambda}(\!\not\!\partial \xi_{\lambda}
-i g [\not\!\!A,\xi_{\lambda}])\psi
$$
and of the forth and fifth terms to
$$
\delta {{\cal L}}_{3/2}(IV+V) =
-\bar{\psi}(\!\not\!\partial \xi_{\lambda}
-i g [\not\!\!A,\xi_{\lambda}])\psi_{\lambda} -
\bar{\psi}_{\lambda}(\!\not\!\partial \xi_{\lambda}
-i g [\not\!\!A,\xi_{\lambda}])\psi
-i g  \bar{\psi}
[\not\!\!A_{\lambda},\xi_{\lambda}])\psi~,
$$
finally the variation of the last term gives
$$
\delta {{\cal L}}_{3/2}(VI) = -{1\over 2}\bar{\psi}(\not\!\partial \xi_{\lambda\lambda}
-i g [\not\!\!A,\xi_{\lambda\lambda}] -2 i  g [\not\!\!A_{\lambda},\xi_{\lambda}])\psi,
$$
therefore the total variation is equal to zero, $\delta {{\cal L}}_{3/2} =0$.

The currents are given by the variation of the action over the tensor gauge fields:
\beqa
J^{a}_{\mu} &=& g\{ \bar{\psi}  \sigma^{a} \gamma_{\mu}  \psi  +
\bar{\psi}_{\lambda}  \sigma^{a} \gamma_{\mu}  \psi_{\lambda}+
{1\over 2}\bar{\psi}_{\lambda\lambda}  \sigma^{a} \gamma_{\mu}  \psi  +
{1\over 2}\bar{\psi}  \sigma^{a} \gamma_{\mu}  \psi_{\lambda\lambda}\},\nn\\
J^{a}_{\mu\nu} &=& g\{ \bar{\psi}  \sigma^{a} \gamma_{\mu}  \psi_{\nu } +
\bar{\psi}_{\nu}  \sigma^{a} \gamma_{\mu}  \psi\},\nn\\
J^{a}_{\mu\lambda\rho} &=& {1\over 2}g
\bar{\psi}  \sigma^{a} \gamma_{\mu}  \psi ~ \eta_{\lambda\rho}.
\eeqa
As one can see from the Lagrangian (\ref{fermionlagrangian})
the interaction of tensor fermions with tensor gauge bosons is going through the
cubic vertex which includes two fermions and a tensor gauge boson, very
similar to the vertices in QED and the Yang-Mills theory.

\section{{\it Interaction of Bosonic Matter and Symmetry Breaking}}

We are in a position now to introduce the gauge invariant interaction of
the tensor gauge bosons with the bosonic matter fields
$\phi_{\lambda_1 ... \lambda_{s}}(x)$. This set of tensor fields
$\{\phi\}$ contains the scalar
field $\phi$ as one of its family members.
The extended isotopic transformation  of the bosonic matter fields
$\phi_{\lambda_1 ... \lambda_{s}}(x)$
we shall define by the formulas \cite{Savvidy:2003fx}
\beqa\label{scalartransformation}
\delta_{\xi}  \phi  &=&-i \tau^{a}\xi^{a}  \phi ,\nonumber\\
\delta_{\xi}  \phi_{\lambda} &=& -i \tau^{a}( \xi^{a} ~\phi_{\lambda}   +
\xi^{a}_{\lambda}~ \phi) ,\nonumber\\
\delta_{\xi}  \phi_{\lambda\rho} &=& -i \tau^{a}( \xi^{a} ~\phi_{\lambda\rho}   +
\xi^{a}_{\lambda}~ \phi_{\rho} + \xi^{a}_{\rho}~ \phi_{\lambda} +
\xi^{a}_{\lambda\rho}~ \phi ),\\
........&.&.......................,\nn
\eeqa
where $\tau^{a}$ are the matrices of the representation $\tau$ of the gauge
group G, according to which the whole family of $\phi's$ transforms.
There is an essential difference
in the transformation properties of the tensor gauge fields
$A_{\mu\lambda_1 ... \lambda_{s}}$
versus
$\phi_{\lambda_1 ... \lambda_{s}}$ . The transformation law for the
bosonic matter fields (\ref{scalartransformation}) is homogeneous,
whereas the transformation of the tensor gauge  fields (\ref{polygauge}),
(\ref{generalgaugetransform}) is inhomogeneous. The general form of the
above transformation is:
\beqa
\delta_{\xi}  \phi_{\lambda_1 ... \lambda_{s}}(x) &=& -i \sum^{s}_{i=0}  \sum_{P's}
 \xi_{\lambda_1 ... \lambda_{i}} ~\phi_{\lambda_{i+1} ... \lambda_{s}}(x), ~~~~~~~
 s=0,1,2,...~~~
\eeqa
and the invariant quadratic form is:
\be\label{invariantform}
I = \sum^{\infty}_{s=0} \Lambda_{s+1} I_s,~~~~~~~I_s = \sum^{2s}_{i=0}~a^{s}_i ~
\phi^{+}_{\lambda_1 ... \lambda_i}~
\phi_{\lambda_{i+1}...\lambda_{2s}}
\sum_{p's} \eta^{\lambda_{i_1} \lambda_{i_2}} .......
\eta^{\lambda_{i_{2s-1}} \lambda_{i_{2s}}}~,
\ee
where the sum $\sum_p$ runs over all permutations of indices
$\lambda_1 ... \lambda_{2s}$ which correspond to nonequal terms and
the numerical coefficient $a^{s}_i = s!/i!(2s-i)!$~. The $\Lambda_s$ (s=1,2,...) are
coupling constants and $\Lambda_1 =1$.
The invariant Lagrangian for scalar field is
$$
{{\cal L}}_{0} =   \nabla^{ij}_{\mu}\phi^{+j} ~ \nabla^{ik}_{\mu}\phi^{k} ~-U(\phi),
$$
where $\nabla^{ij}_{\mu} = \delta^{ij} \partial_{\mu} -ig \tau^{ij}_{a} A^{a}_{\mu}$.
The covariant derivatives for the low-rank tensors transform as follows:
\beqa\label{scalartransformationcovariant}
\delta_{\xi}  \nabla_{\mu}\phi  &=&-i \xi   \nabla_{\mu} \phi , \\
\delta_{\xi}   \nabla_{\mu} \phi_{\lambda} &=& -i\xi  \nabla_{\mu} \phi_{\lambda} -i
\xi_{\lambda} \nabla_{\mu} \phi -i \nabla_{\mu}\xi_{\lambda}~  \phi ,\nonumber\\
\delta_{\xi}  \nabla_{\mu}\phi_{\lambda\rho} &=& -i( \xi \nabla_{\mu} \phi_{\lambda\rho}   +
\xi_{\lambda}\nabla_{\mu}\phi_{\rho} + \xi_{\rho}\nabla_{\mu} \phi_{\lambda} +
\xi_{\lambda\rho}\nabla_{\mu}\phi )\nn\\
&~&-i (\nabla_{\mu} \xi_{\lambda}\phi_{\rho}+
\nabla_{\mu} \xi_{\rho}\phi_{\lambda}+\nabla_{\mu} \xi_{\lambda\rho}\phi).\nn
\eeqa
The invariant Lagrangian will take the form:
\beqa\label{scalarlagrangian}
{{\cal L}}_{\phi} &=&
\nabla_{\mu}\phi^{+} ~ \nabla_{\mu}\phi + b_2 \{
\nabla_{\mu}\phi^{+}_{\lambda} ~\nabla_{\mu}\phi_{\lambda} +
{1\over 2}\nabla_{\mu}\phi^{+}_{\lambda\lambda} ~ \nabla_{\mu}\phi +
{1\over 2}\nabla_{\mu}\phi^{+} ~ \nabla_{\mu}\phi_{\lambda\lambda} +\nonumber\\
&-& ig \nabla_{\mu}\phi^{+} ~A_{\mu\lambda}\phi_{\lambda} +
ig \phi^{+}_{\lambda} A_{\mu\lambda}~\nabla_{\mu}\phi -
ig \nabla_{\mu}\phi^{+}_{\lambda} ~A_{\mu\lambda}\phi+
ig \phi^{+} A_{\mu\lambda}  ~ \nabla_{\mu}\phi_{\lambda}-\nn\\
&+&g^2  \phi^{+} A_{\mu\lambda} A_{\mu\lambda}\phi-
{1\over 2}ig \nabla_{\mu}\phi^{+} ~A_{\mu\lambda\lambda}\phi +
{1\over 2}ig \phi^{+} A_{\mu\lambda\lambda}~\nabla_{\mu}\phi \}~-~U(\phi),
\eeqa
where $b_2$ is the new coupling constant.
The variation of the Lagrangian has the following terms:
$$
\delta {{\cal L}}_{\phi}(II) = i \nabla_{\mu}\phi^{+}\xi_{\lambda}\nabla_{\mu}\phi_{\lambda}
-i \nabla_{\mu}\phi^{+}_{\lambda}\xi_{\lambda}\nabla_{\mu}\phi
+i \phi^{+}\nabla_{\mu}\xi_{\lambda} \nabla_{\mu}\phi_{\lambda}
-i \nabla_{\mu}\phi^{+}_{\lambda}\nabla_{\mu}\xi_{\lambda}\phi,
$$
\beqa
\delta {{\cal L}}_{\phi}(III + IV) &=&
i \nabla_{\mu}\phi^{+}_{\lambda}\xi_{\lambda}\nabla_{\mu}\phi -
i \nabla_{\mu}\phi^{+}\xi_{\lambda}\nabla_{\mu}\phi_{\lambda}
+i \phi^{+}_{\lambda}\nabla_{\mu}\xi_{\lambda} \nabla_{\mu}\phi
-i \nabla_{\mu}\phi^{+}\nabla_{\mu}\xi_{\lambda}\phi_{\lambda}\nn\\
&+&{i\over 2} \phi^{+}\nabla_{\mu}\xi_{\lambda\lambda} \nabla_{\mu}\phi
-{i\over 2} \nabla_{\mu}\phi^{+}\nabla_{\mu}\xi_{\lambda\lambda}\phi,\nn
\eeqa
\beqa
\delta {{\cal L}}_{\phi}(V +VI) =
i \nabla_{\mu}\phi^{+}\nabla_{\mu}\xi_{\lambda}\phi_{\lambda}
-i \phi^{+}_{\lambda}\nabla_{\mu}\xi_{\lambda} \nabla_{\mu}\phi
-g \nabla_{\mu}\phi^{+} A_{\mu\lambda}\xi_{\lambda}\phi
-g \phi^{+}\xi_{\lambda}A_{\mu\lambda}\nabla_{\mu}\phi,\nn
\eeqa
\beqa
\delta {{\cal L}}_{\phi}(VII +VIII) &=&
g \phi^{+}A_{\mu\lambda}\xi_{\lambda}\nabla_{\mu}\phi
+g \nabla_{\mu}\phi^{+}\xi_{\lambda} A_{\mu\lambda}\phi
+i \nabla_{\mu}\phi^{+}_{\lambda} \nabla_{\mu}\xi_{\lambda}\phi
-i \phi^{+}\nabla_{\mu}\xi_{\lambda}\nabla_{\mu}\phi_{\lambda}\nn\\
&+&g \phi^{+}\nabla_{\mu}\xi_{\lambda} A_{\mu\lambda}\phi
+g \phi^{+} A_{\mu\lambda}\nabla_{\mu}\xi_{\lambda}\phi~,
\eeqa
together with the ninth term
$$
\delta {{\cal L}}_{\phi}(IX) = -g \phi^{+}\nabla_{\mu}\xi_{\lambda} A_{\mu\lambda}\phi
-g \phi^{+} A_{\mu\lambda}\nabla_{\mu}\xi_{\lambda}\phi ,
$$
and finally the variation of the last two terms gives
$$
\delta {{\cal L}}_{\phi}(X + XI) = {i\over 2}\nabla_{\mu}\phi^{+}
(\nabla_{\mu}\xi_{\lambda\lambda}- 2 i g [A_{\mu\lambda} \xi_{\lambda}])\phi
-{i\over 2}\phi^{+}(\nabla_{\mu}\xi_{\lambda\lambda}
- 2 i g [A_{\mu\lambda} \xi_{\lambda}])\nabla_{\mu}\phi,
$$
therefore the total variation is equal to zero, $\delta {{\cal L}}_{\phi} =0$.

We have to introduce the invariant self-interaction Lagrangian for the extended scalar
sector. The quadratic form which is invariant with respect to the extended
homogeneous transformations (\ref{scalartransformation}) has the form
(\ref{invariantform})
\be\label{invariantquadraticform}
I= \phi^{\dag} \phi +
\Lambda_2 (\phi^{\dag}_{\lambda}\phi_{\lambda} + {1\over 2}
\phi^{\dag}\phi_{\lambda\lambda}  +{1\over 2}
\phi^{\dag}_{\lambda\lambda}  \phi).
\ee
Its invariance can be confirmed by direct calculation similar to the one we
performed above. Using this quadratic form we can construct the invariant
potential as
\be
U(\phi) ={1\over 2} \lambda^2 [\phi^{\dag} \phi +
\Lambda_2(\phi^{\dag}_{\lambda}\phi_{\lambda} + {1\over 2}
\phi^{\dag}\phi_{\lambda\lambda}  +{1\over 2}
\phi^{\dag}_{\lambda\lambda}  \phi) - {1\over 2} \eta^2]^2 ~
\ee
so that the vacuum expectation value of the scalar field will be as in the
standard model:
$$
\phi_{vac} = \eta /\sqrt{2}.
$$
As one can see, the vacuum expectation value of the non-gauge bosons is equal to zero
and does not break the Poincar\'e invariance.  The Higgs boson mass therefore
remains the same as in the standard model:
$$
m_H = \lambda \eta.
$$
The new non-gauge vector boson $\phi_{\lambda}$,  which is predicted by our model,
will also acquire mass through the interaction term
\be
{1\over 2} ~\lambda^2 ~2~\Lambda_2~ \phi^{\dag} \phi ~
\phi^{\dag}_{\lambda}\phi_{\lambda}~~~
\rightarrow ~~~\lambda^2 ~\Lambda_2~
<\phi^{\dag}> <\phi> \phi^{\dag}_{\lambda}\phi_{\lambda}=
~\Lambda_2~{\lambda^2 \eta^2 \over 2} \phi^{\dag}_{\lambda}\phi_{\lambda}
\ee
and is equal to the mass of the standard Higgs scalar:
\be
m_{\phi} = \sqrt{{\Lambda_2 \over 2 b_2}} \lambda \eta  =
\sqrt{{\Lambda_2 \over 2 b_2}} m_H~.
\ee
From the Lagrangian (\ref{scalarlagrangian}) (the first term in the third line)
we can also read off the induced mass of the tensor gauge bosons:
\be
M_{ab} = ({b_2 \over g_2})~2 g^2  <\phi^{\dag}\tau_{a} ><\tau_{b} \phi>,
\ee
when the scalar field $\phi$ gets non-zero vacuum expectation value. Thus the
masses of the tensor gauge bosons are
\be
m^{2}_{T} = ({b_2 \over g_2}) m^{2}_V.
\ee
We conclude that the proposed extension of the non-Abelian gauge theory,
which interacts with the tensor matter, predicts
at the tree level degeneracy of the mass spectrum of the new tensor gauge
bosons.

It is not completely clear to the author, whether the number
of gauge parameters $\xi, \xi_\lambda ,...$ is
sufficient to exclude simultaneously not only negative norm states
of the tensor gauge bosons, but also negative norm
states of the bosonic matter fields of non-gauge
nature  $\phi_{\lambda}, \phi_{\lambda\rho},...$. This question requires detailed
analysis and we shall leave this question for the future studies.

\section{\it Tensor Extension of  Electroweak Theory}

Let us consider the possible extension of the standard model of electroweak
interactions which follows from the above generalization. In the first model
which we shall consider only the $SU(2)_L$ group will be extended to higher spins,
but not the $U(1)_Y$ group.  The  $W^{\pm},Z$  gauge bosons will
receive their higher-spin descendence
\beqa
&~&(W^{\pm},Z)_{\mu},~~(\tilde{W}^{\pm},\tilde{Z})_{\mu\lambda}, .....
\eeqa
and the doublet of complex Higgs
scalars will appear together with their higher-spin partners:
$$
(\begin{array}{c}
  \phi^+ \\
   \phi^o
\end{array})~~, ~~~~~~~(\begin{array}{c}
  \phi^+ \\
   \phi^o
\end{array})_{\lambda}~~, ~~~~~~~(\begin{array}{c}
  \phi^+ \\
   \phi^o
\end{array})_{\lambda\rho}~~,......~~~~~~~~~~~~~~~Y=+1.
$$
The Lagrangian which describes  the interaction of the tensor gauge bosons with
scalar fields and tensor bosons is:
\beqa\label{standardmodellagrangian}
{{\cal L}}  =&-&{1\over 4}G^{i}_{\mu\nu}
G^{i}_{\mu\nu} - {1\over 4}F_{\mu\nu}
F_{\mu\nu} - (\partial_{\mu}   + {ig^{'} \over 2} B_{\mu} +
{ig \over 2}\tau^{i} A^{i}_{\mu} )\phi^{\dag} ~
(\partial_{\mu}   - {ig^{'} \over 2}B_{\mu} -
{ig \over 2}\tau^{i}A^{i}_{\mu} )\phi +\nn\\
&+&g_2\{-{1\over 4}G^{i}_{\mu\nu,\lambda}G^{i}_{\mu\nu,\lambda}
-{1\over 4}G^{i}_{\mu\nu}G^{i}_{\mu\nu,\lambda\lambda}\} -
b_2 \{{g^{2} \over 4} \phi^{\dag} \tau^{i} A^{i}_{\mu\lambda} \tau^{j} A^{j}_{\mu\lambda}\phi \nn\\
&+&\nabla_{\mu}\phi^{\dag}_{\lambda} ~\nabla_{\mu}\phi_{\lambda} +
{1\over 2}\nabla_{\mu}\phi^{\dag}_{\lambda\lambda} ~ \nabla_{\mu}\phi +
{1\over 2}\nabla_{\mu}\phi^{\dag} ~ \nabla_{\mu}\phi_{\lambda\lambda} \nonumber\\
&-& ig \nabla_{\mu}\phi^{\dag} ~A_{\mu\lambda}\phi_{\lambda} +
ig \phi^{\dag}_{\lambda} A_{\mu\lambda}~\nabla_{\mu}\phi -
ig \nabla_{\mu}\phi^{\dag}_{\lambda} ~A_{\mu\lambda}\phi+
ig \phi^{\dag} A_{\mu\lambda}  ~ \nabla_{\mu}\phi_{\lambda}-\nn\\
&-&{1\over 2}ig \nabla_{\mu}\phi^{\dag} ~A_{\mu\lambda\lambda}\phi +
{1\over 2}ig \phi^{\dag} A_{\mu\lambda\lambda}~\nabla_{\mu}\phi ~\} - U(\phi),
\eeqa
where
$$
\nabla_{\mu}  = \partial_{\mu}   - {ig^{'} \over 2} Y B_{\mu} - ig  T^{i} A^{i}_{\mu},
$$
Y  is hypercharge, Q is charge, $Q = T_3 + Y/2$, and for isospinor fields
$T^i = \tau^i /2 $. The three terms in the first line represent the
standard electroweak model and the rest
of the terms - its higher-spin generalization. Therefore all parameters of the standard
model are incorporated in the extension.
The third  term in the second line will
generate the masses of the tensor $\tilde{W}^{\pm},\tilde{Z}$  gauge bosons:
\be
{1\over 8} ({b_2 \over g_2}) g^2 \eta^2 [(A^{3}_{\mu\lambda})^2 +
2 A^{+}_{\mu\lambda}A^{-}_{\mu\lambda}],
\ee
when the scalar fields acquire  the vacuum expectation value $\eta$:
$$
\phi = {1\over \sqrt{2}} (\begin{array}{c}
                           0 \\
                           \eta + \chi(x)
                         \end{array})
$$
and
$$
\tilde{Z}_{\mu\lambda} =A^{3}_{\mu\lambda},~~~   \tilde{W}^{\pm}_{\mu\lambda}=
{1\over \sqrt{2}}(A^{1}_{\mu\lambda} \pm  i A^{2}_{\mu\lambda}),
$$
Thus all intermediate spin-2 bosons will acquire the same mass
\be
m^{2}_{\tilde{W},\tilde{Z}} = ({b_2 \over g_2}) m^{2}_W~~.
\ee
The rest of the
terms describe the interaction between "old" and new particles.
One should also introduce
the Yukawa self-interaction  for the bosons $\phi_{\lambda}$
in order to make them massive.

Finally let us consider the fermion sector of the extended electroweak model.
One should note that the interaction of tensor gauge bosons with fermions is not as
usual as one could expect. Indeed, let us now analyze the interaction with new
spinor-tensor leptons
$$
L = {1\over 2}(1+ \gamma_{5})(\begin{array}{c}
  \nu_e \\
   e
\end{array}),~~~~L_{\lambda} =
{1\over 2}(1+ \gamma_{5})(\begin{array}{c}
  \nu_e \\
   e
\end{array})_{\lambda},~~~~L_{\lambda\rho} =
{1\over 2}(1+ \gamma_{5})(\begin{array}{c}
  \nu_e \\
   e
\end{array})_{\lambda\rho}~..~~~Y=-1.
$$
All these left-handed states have hypercharge $Y=-1$ and the only right-handed
state
$$
R = {1\over 2}(1 - \gamma_{5})e, ~~~~~~~~~~~~~~~~~~~~Y=-2
$$
has the hypercharge $Y = -2$. The corresponding Lagrangian will take the form
\beqa
{{\cal L}}_{F}&=& \bar{L} \not\!\nabla L + \bar{R} \not\!\nabla R  +\\
&+&f_1 ~\{\bar{L}_\lambda \not\!\nabla L_\lambda + {1\over 2}\bar{L} \not\!\nabla L_{\lambda\lambda} +
{1\over 2}\bar{L}_{\lambda\lambda} \not\!\nabla L +\nn\\
&+&g \bar{L}_\lambda \not\!\!A_{\lambda}  L  + g \bar{L} \not\!\!A_{\lambda}  L_\lambda
+ {1\over 2}g \bar{L} \not\!\!A_{\lambda\lambda}  L  \}~,   \nn
\eeqa
where the first two terms describe the standard electroweak interaction of
vector gauge bosons with standard spin-1/2 leptons,
the next three terms describe the interaction of the vector gauge bosons
with new leptons of the spin 3/2 and finally the last three terms describe
the interaction of the new tensor gauge bosons
$\tilde{W}^{\pm},\tilde{Z}$ with standard spin-1/2 and
spin-3/2 leptons.

The new interaction vertices generate
decay of the standard vector gauge bosons through the channels
$$
\gamma,Z \rightarrow e_{3/2}+\bar{e}_{3/2},~~~
\gamma,Z \rightarrow \nu_{3/2}+\bar{\nu}_{3/2},~~~
W \rightarrow \nu_{3/2} + e_{3/2},~~~W \rightarrow ~\nu_{3/2}+e_{3/2}~,
$$
where a pair of new leptons is created.
The observability of these channels depends on the masses of the new
leptons. This information is encoded into the Yukawa couplings, as it takes place
for the standard leptons of the spin 1/2. We can only say that they are large
enough not to be seen at low energies, but are predicted to be visible
at higher-energy experiments.

The decay reactions of the new tensor gauge bosons $\tilde{W}^{\pm},\tilde{Z}$
can take place through the channels
\be\label{decaychannels}
\tilde{Z}\rightarrow e_{3/2}+\bar{e}_{1/2},~~~
\tilde{Z}\rightarrow \nu_{3/2}+\bar{\nu}_{1/2},~~~
\tilde{W} \rightarrow ~\nu_{1/2}+e_{3/2},~~~
\tilde{W} \rightarrow ~\nu_{3/2}+e_{1/2}.
\ee
The main feature of these processes is that they create a pair which
consists of a standard lepton $e_{1/2}$ and of a new lepton $e_{3/2}$
of the spin 3/2. Because in all these
reactions there always participates a new lepton, they may
take place also at large enough energies, but it is impossible
to predict the threshold energy
because we do not know the corresponding Yukawa couplings. The
situation with Yukawa couplings is the same as it is in the standard model.
There is no decay channels of the new tensor bosons only into the standard leptons,
as one can see from the Lagrangian. Therefore it is also impossible to create
tensor gauge bosons directly in $e^+  + e^-$ annihilation, but they can appear in the
decay of the  $Z $
\be\label{mostpromising}
e^+  + e^- \rightarrow Z \rightarrow \tilde{W}^+ + \tilde{W}^- ~~
\ee
and will afterwards decay through the channels discussed
above (\ref{decaychannels}) $\tilde{W} \rightarrow ~\nu + \tilde{e}$ or
$\tilde{W} \rightarrow ~\tilde{\nu} + e$ .
{\it It seems that reaction (\ref{mostpromising}), predicted by the generalized
theory, is the most appropriate candidate which could be tested in
the experiment}. The details will be given in the forthcoming publication.

We did not consider the tensor extension of the $U(1)_Y$ in the first place,
because in that case we
shall have the massless spin-2 descendent of the photon, the existence of which
most probably will be ruled out by experiment. But, in principle, the models
of this type can be considered and it is worth to study them. The right-handed
sector should be enlarged in the following way:
$$
R = {1\over 2}(1 - \gamma_{5})e, ~~~~R_{\lambda} = {1\over 2}(1 - \gamma_{5})e_{\lambda}, ~~~~
R_{\lambda\rho} = {1\over 2}(1 - \gamma_{5})e_{\lambda\rho},~~...,~~~~~~~~~~~~~~~Y=-2,
$$
and the Lagrangian will take the form
\beqa
{{\cal L}}_{F}&=& \bar{L} \not\!\nabla L + \bar{R} \not\!\nabla R \\
&+&f_1\{\bar{L}_\lambda \not\!\nabla L_\lambda + {1\over 2}\bar{L} \not\!\nabla L_{\lambda\lambda} +
{1\over 2}\bar{L}_{\lambda\lambda} \not\!\nabla L +
g \bar{L}_\lambda \not\!\!A_{\lambda}  L  + g \bar{L} \not\!\!A_{\lambda}  L_\lambda
+ {1\over 2}g \bar{L} \not\!\!A_{\lambda\lambda}  L     \nn\\
&+&\bar{R}_\lambda \not\!\nabla R_\lambda + {1\over 2}\bar{R} \not\!\nabla R_{\lambda\lambda} +
{1\over 2}\bar{R}_{\lambda\lambda} \not\!\nabla R +
g^{'} \bar{R}_\lambda \not\!\!B_{\lambda}  R  + g^{'} \bar{R} \not\!\!B_{\lambda}  R_\lambda
+ {1\over 2}g^{'} \bar{R} \not\!\!B_{\lambda\lambda}  R\},\nn
\eeqa
where the terms in the last line describe the interaction of the Abelian $U(1)_Y$
tensor fields $B_{\mu},B_{\mu\lambda},... $
with the right-handed sector of new leptons.

\section{\it Tensor Extension of QCD}

It is also appealing to consider the tensor extension of QCD. The gauge
group $SU(3)_c$ can be extended in the way described above and should
contain additional tensor gluons
$$
A_{\mu},~~A_{\mu\lambda}, .....
$$
together with the higher-spin quarks
$$
q,~~q_{\lambda}, .....
$$
The physical consequences and the phenomena of the
gauge field condensation predicted in \cite{Savvidy:1977as} will be studied
in the extended model in the forthcoming publication.
One should stress again that within the given extension it is impossible
to predict masses of the tensor quarks, as it has been impossible to predict them
within the standard model.

\section{{\it Symmetrized Extended Gauge Transformation}}

Let us consider the situation when the
higher-rank tensor fields $A_{\lambda_1 ...\lambda_s}$ are totally symmetric tensors.
In that case the gauge transformation (\ref{polygauge}), (\ref{matrixform}),
(\ref{generalgaugetransform}) should be modified in order
to respect the symmetry properties of the tensor fields.
Therefore we shall symmetrize the extended gauge transformation (\ref{polygauge}),
(\ref{generalgaugetransform}) over all indices as follows:
\beqa\label{polygaugesymmetric}
\tilde{\delta} A^{a}_{\mu} &=& ( \delta^{ab}\partial_{\mu}
+g f^{acb}A^{c}_{\mu})\xi^b ,~~~~~ \nonumber\\
\tilde{\delta} A^{a}_{\mu\nu} &=& \underline{( \delta^{ab}\partial_{\mu}
+  g f^{acb}A^{c}_{\mu})\xi^{b}_{\nu} }+ g f^{acb}A^{c}_{\mu\nu}\xi^{b},\nonumber\\
\tilde{\delta} A^{a}_{\mu\nu \lambda}& =& \underline{( \delta^{ab}\partial_{\mu}
+g f^{acb} A^{c}_{\mu})\xi^{b}_{\nu\lambda}} +
g f^{acb}(\underline{A^{c}_{\mu  \nu}\xi^{b}_{\lambda }} +A^{c}_{\mu\nu\lambda}\xi^{b}),\\
........&.&......................................,\nn
\eeqa
where we explicitly symmetrize the right-hand side of
these equations over all space-time indices:
\beqa
\underline{( \delta^{ab}\partial_{\mu}
+  g f^{acb}A^{c}_{\mu})\xi^{b}_{\nu} } &=& ( \delta^{ab}\partial_{\mu}
+  g f^{acb}A^{c}_{\mu})\xi^{b}_{\nu} + ( \delta^{ab}\partial_{\nu}
+  g f^{acb}A^{c}_{\mu})\xi^{b}_{\mu}\nn\\
\underline{A^{c}_{\mu  \nu}\xi^{b}_{\lambda }} &=&
A^{c}_{\mu  \nu}\xi^{b}_{\lambda } +A^{c}_{\nu \lambda}\xi^{b}_{\mu  }
+A^{c}_{ \lambda  \mu}\xi^{b}_{\nu}\nn
\eeqa
and so on. One should also require that the
gauge parameters $\xi^{a}_{\lambda_1 ...\lambda_s}$ are totally symmetric
tensors. For example, in a matrix notation, we have
\beqa
\tilde{\delta}_{\xi}  A_{\mu\nu} = \partial_{\mu}\xi_{\nu} -i g[A_{\mu},\xi_{\nu}]
+\partial_{\nu}\xi_{\mu}  -i g[A_{\nu},\xi_{\mu}] -i g [A_{\mu\nu},\xi] .\nonumber
\eeqa
First let us check, whether these transformations still form a closed algebra.
The commutator of the above gauge transformations acting on rank-2 tensor is:
\beqa
[\tilde{\delta}_{\eta},\tilde{\delta}_{\xi}]A_{\mu\nu} &=& -i g~ \delta_{\eta}~(-i g[A_{\mu},\xi_{\nu}]
-i g[A_{\nu},\xi_{\mu}]-i g [A_{\mu\nu},\xi])+\nonumber\\
&~&~~ i g~\delta_{\xi}~(-i g[A_{\mu},\eta_{\nu}]
-i g[A_{\nu},\eta_{\mu}] -i g [A_{\mu\nu},\eta])\nonumber\\
&=&-ig \{~\partial_{\mu}([\eta,\xi_{\nu}] +[\eta_{\nu},\xi]) -i
g[A_{\mu},[\eta,\xi_{\nu}] +[\eta_{\nu},\xi]] +\nonumber\\
&~&\partial_{\nu}([\eta,\xi_{\mu}] +[\eta_{\mu},\xi]) -i
g[A_{\nu},[\eta,\xi_{\mu}] +[\eta_{\mu},\xi]] -i
g[A_{\mu\nu},[\eta,\xi]] \}\nonumber
\eeqa
and is again a gauge transformation with gauge parameters
$$
\zeta =[\eta,\xi],~~~~~~~\zeta_{\mu} = [\eta,\xi_{\nu}] +[\eta_{\nu},\xi].
$$
We arrive to the same conclusion calculating the commutator of two
gauge transformations acting on a third-rank tensor field:
\be
[\tilde{\delta}_{\eta},\tilde{\delta}_{\xi}]A_{\mu\nu\lambda} = -i g~
\tilde{\delta}_{\zeta} A_{\mu\nu\lambda},
\ee
where the gauge parameters form the same algebra as before (\ref{commutatorofparameters}).
Therefore we conclude that the symmetrized gauge transformations
still form a closed algebraic structure.

The gauge transformation of the field strength $G^{a}_{\mu\nu,\lambda}$ will take the
form
\beqa\label{inhomogeneous}
\tilde{\delta} G^{a}_{\mu\nu,\lambda} &=& g f^{abc} (~G^{b}_{\mu\nu,\lambda} \xi^c
+ G^{b}_{\mu\nu} \xi^{c}_{\lambda}~) +
\nabla^{ab}_{\mu} \nabla^{bc}_{\lambda}\xi^{c}_{\nu} -
\nabla^{ab}_{\nu} \nabla^{bc}_{\lambda}\xi^{c}_{\mu}.
\eeqa
This transformation is inhomogeneous and is essentially different from
the extended gauge transformation
of the field strength considered in the main text (\ref{fieldstrengh3thtransfor}):
$$
\delta G^{a}_{\mu\nu,\lambda}=g f^{abc} (~G^{b}_{\mu\nu,\lambda} \xi^c
+ G^{b}_{\mu\nu} \xi^{c}_{\lambda}~).
$$
The additional two terms,
which break the homogeneity of the gauge transformation of the
field strength,
appear because we have introduced the symmetrization with respect to
all indices in the extended gauge transformation (\ref{polygaugesymmetric}).
These inhomogeneous terms and the appearance of non-commutative covariant derivatives
in the field strength-curvature transformation
(\ref{inhomogeneous}) are the main
obstacles in the construction of the interacting field theory with
totally symmetric tensors.

In the weak coupling limit $g \rightarrow 0$ the transformation
(\ref{polygaugesymmetric}) reduces to the following form:
\beqa\label{symmetricgaugetransformation}
\tilde{\delta}_{\xi}  A_{\mu} &=& \partial_{\mu}\xi  \nonumber\\
\tilde{\delta}_{\xi}  A_{\mu\nu} &=& \partial_{\mu} \xi_{\nu} +
\partial_{\nu} \xi_{\mu},\nonumber\\
\tilde{\delta}_{\xi}  A_{\mu\nu\lambda} &=& \partial_{\mu}\xi_{\nu\lambda} +
 \partial_{\nu}\xi_{\lambda\mu}  +  \partial_{\lambda}\xi_{\mu\nu},
\eeqa
and coincides with the transformation considered by Fronsdal \cite{fronsdal}
and other authors \cite{fierzpauli,Gupta,kraichnan,thirring,feynman,deser,fronsdal2}.
The above transformation was considered as a zero-order approximation for some yet
unknown transformation, which should appear as its higher-order deformation.
The above transformation law $\tilde{\delta}_{\xi}$
suggests a particular solution and partially explains why this
approach leads to difficulties.

\section{\it Conclusion}

In the present paper we have extended the gauge principle so that it enlarges
the original algebra of the Abelian local gauge transformations
found in \cite{Savvidy:2003fx} to the non-Abelian case
and unify into one multiplet particles with arbitrary spins.
As we have seen, this leads to a natural generalization of the Yang-Mills theory
which describes interaction of tensor gauge bosons.
The key ingredient of the extension was a previous
discovery of the  Abelian higher-spin gauge transformations of the
ground state wave function of the tensionless string theory
\cite{Savvidy:2003fx} (expression (64) in \cite{Savvidy:2003fx}).
The extended non-Abelian gauge transformations
defined for the tensor gauge fields have led us to the construction of the
appropriate field strength tensors and of the invariant Lagrangian.

As an example of an extended gauge field theory with infinite many gauge fields,
this theory {\it can be associated} with
the field theory of the tensionless strings, because in
our generalization of the non-Abelian Yang-Mills theory we essentially
used the symmetry group which appears as symmetry of the ground state
wave function of the tensionless strings. Nevertheless I do not know how to
derive it directly from tensionless strings.

The proposed extension may lead  to a natural inclusion of the standard
theory of fundamental forces into a larger theory in which standard
particles (vector gauge bosons, leptons and quarks) represent a
low-spin subgroup of an enlarged family of particles with higher spins.
The conjectured extension of the fundamental forces can in principle be
checked in future experiments.

There are some interesting observations which follow from the  proposed theory
concerning the structure of elementary particles and their constituents.
We do not know, whether the proposed theory indeed describes the observed
micro-cosmos, but if it does, then it might suggest some new insight
to the fundamental questions of the atomic theory,
that is, whether there exist the most elementary constituents of the matter in
micro-cosmos - its ultimate building blocks, and
if indeed they exist, then how many they are.
As one can see, if one supposes that the ultimate building blocks are the gauge
bosons, quarks and leptons, then
there is a possibility that they are just the first members
of the constituent structure with infinite many tensor particles.

I would like to thank Konstantin Savvidy for collaboration
in the field of tensionless strings.
The discussions and support of Lars Brink and Ludwig Faddeev are greatly appreciated.
I would like to express my gratitude to Jan Ambjorn and Des Johnson for their
long term support.
I wish also to thank  Luis  Alvarez-Gaume, Ignatios  Antoniadis, Ioannis Bakas,
Raymond Stora and Peter Minkowski  for discussions and CERN Theory
Division, where part of this work was completed, for hospitality.
This work was partially supported by the EEC Grant no. HPRN-CT-1999-00161 and
EEC Grant no. MRTN-CT-2004-005616.

\section{{\it Appendix}}

In order to prove that the extended gauge
transformations form a closed algebraic structure in general case one
should us the following identities:
\be
\sum^{s}_{i=0}  \sum_{P's} [~\eta_{\lambda_{i+1} ...\lambda_s},
\partial_{\mu}\xi_{\lambda_{1} ...\lambda_i }~]=\sum^{s}_{i=0}  \sum_{P's}
[~\eta_{\lambda_1 ...\lambda_i},
\partial_{\mu}\xi_{\lambda_{i+1} ...\lambda_s }~],
\ee
\be
\sum^{s}_{i=0}  \sum_{P's} [[~A_{\mu},\xi_{\lambda_1 ...\lambda_i}],
\eta_{\lambda_{i+1} ...\lambda_s }~] =\sum^{s}_{i=0}  \sum_{P's}
[[A_{\mu},\xi_{\lambda_{i+1} ...\lambda_s }],\eta_{\lambda_1 ...\lambda_i}~],
\ee
\be
\sum^{i}_{j=1}  \sum_{P's} [[~A_{\mu\lambda_1 ...\lambda_j},
\xi_{\lambda_{j+1} ...\lambda_i}],
\eta_{\lambda_{i+1} ...\lambda_s }~] =\sum^{i}_{j=1}  \sum_{P's}
[[A_{\mu\lambda_1 ...\lambda_j},\xi_{\lambda_{i+1} ...\lambda_s }],
\eta_{\lambda_{j+1} ...\lambda_i}~].
\ee
In particular, they allow to demonstrate that
$$
\sum^{s}_{i=0}  \sum_{P's} [~\partial_{\mu}\eta_{\lambda_1 ...\lambda_i},
\xi_{\lambda_{i+1} ...\lambda_s }~]+[~\eta_{\lambda_{i+1} ...\lambda_s},
\partial_{\mu}\xi_{\lambda_{1} ...\lambda_i }~]=\sum^{s}_{i=0}  \sum_{P's}
\partial_{\mu}[~\eta_{\lambda_1 ...\lambda_i},
\xi_{\lambda_{i+1} ...\lambda_s }~].
$$
One should also use the Jacoby identity in the form
$$
[[A_{\mu\lambda_1 ...\lambda_j},\eta_{\lambda_{j+1} ...\lambda_i}],
\xi_{\lambda_{i+1} ...\lambda_s }] +
[[\xi_{\lambda_{i+1} ...\lambda_s },A_{\mu\lambda_1 ...\lambda_j}],
\eta_{\lambda_{j+1} ...\lambda_i}] =
[A_{\mu\lambda_1 ...\lambda_j},[\eta_{\lambda_{i+1} ...\lambda_s },
\xi_{\lambda_{j+1} ...\lambda_i}]].
$$

\vfill
\end{document}